%% file: arxiv.tex
\documentclass{article}
\usepackage[utf8]{inputenc}
\usepackage{amssymb}
\usepackage{geometry}
\usepackage[usenames,dvipsnames]{color}
\usepackage{bbold}
\usepackage{slashed}
\usepackage[table]{xcolor}
\usepackage{graphicx}
\usepackage{mathtools}
\usepackage{svg}
\usepackage{adjustbox}
\svgpath{{./Images/}} 

\usepackage{todonotes}
\usepackage{comment}
\usepackage{subfloat}
\usepackage{subcaption}
\usepackage{tabularx}
\usepackage{float}
\usepackage{wrapfig}
\usepackage{booktabs}
\usepackage{hyperref}
\usepackage{amsthm}
\usepackage{enumitem}
\usepackage{bm}
\usepackage{dsfont}

\usetikzlibrary{quantikz}
\usetikzlibrary{patterns}
\usepackage{pgfplots}
\usepgfplotslibrary{groupplots}
\usepgfplotslibrary{fillbetween}
\usepackage{tikz-qtree}

\usepackage[
backend=biber,
style=alphabetic,
sorting=ynt
]{biblatex}

\usepackage{multicol}
\usepackage{algorithm,algpseudocode}
\algdef{SE}[SUBALG]{Indent}{EndIndent}{}{\algorithmicend\ }%
\algtext*{Indent}
\algtext*{EndIndent}
\renewcommand{\Comment}[2][.5\linewidth]{%
  \leavevmode\hfill\makebox[#1][l]{//~#2}}
  
\newtheorem{theorem}{Theorem}[section]
\newtheorem{lemma}{Lemma}
\newtheorem{remark}{Remark}[section]
\renewenvironment{proof}[1][\proofname]{{\itshape \bfseries #1. }}{\qed}

\definecolor{rosaUDC}{RGB}{210, 0, 123}
\definecolor{mylightyellow}{rgb}{1,1,.8}
\definecolor{mylightgreen}{rgb}{.8,1,.8}
\definecolor{mydarkred}{RGB}{178,34,34}
\definecolor{mydarkgreen}{RGB}{34,139,34}
\definecolor{mydarkblue}{RGB}{72,61,139}
\definecolor{mydarkyellow}{RGB}{218,165,32}
\definecolor{g}{rgb}{.8,.8,.8}

\newcommand{\norm}[1]{\left\lVert#1\right\rVert}
\newcommand{\abs}[1]{\left\lvert#1\right\rvert}

\usepackage{setspace}

\newcounter{mycomment}

\graphicspath{{./images/}}
\title{Real Option Pricing using Quantum Computers}
\author{Alberto Manzano$^{a}$,
Gonzalo Ferro Costas$^{b}$,
Álvaro Leitao$^{a}$,\\
Carlos Vázquez$^{a}$ and Andrés Gómez$^{b}$.
}

\date{}

\begin{document}

\maketitle
\begin{center}\it{
$^{a}$Department of Mathematics and CITIC, Universidade da Coruña, A Coruña, Spain\\
\vspace{15pt}
$^{b}$Centro de Supercomputación de Galicia (CESGA), Santiago de Compostela, Spain
}
\end{center}
\vspace{25pt}
\input{body/abstract}
\newpage

\tableofcontents
\newpage

\input{body/body}

\printbibliography

\newpage
\appendix

\input{body/appendix1}
\input{body/appendix2}

\end{document}

%% file: body/abstract.tex
\abstract{In this work we present an alternative methodology to the standard Quantum Accelerated Monte Carlo (QAMC) applied to derivatives pricing. Our pipeline benefits from the combination of a new encoding protocol, referred to as the direct encoding, and a amplitude estimation algorithm, the modified Real Quantum Amplitude Estimation (mRQAE) algorithm. On the one hand, the direct encoding prepares a quantum state which contains the information about the sign of the expected payoff. On the other hand, the mRQAE is able to read all the information contained in the quantum state. Although the procedure we describe is different from the standard one, the main building blocks are almost the same. Thus, all the extensive research that has been performed is still applicable. Moreover, we experimentally compare the performance of the proposed methodology against the standard QAMC employing a quantum emulator and show that we retain the speedups.}

%% file: body/body.tex
 \section{Introduction}\label{sec:introduction}

 Over the last few years there has been an increasing interest in the application of quantum computing to quantitative finance. One of the reasons for such boost of popularity
is because many algorithms used by financial institutions demand a high computing
capacity and quantum computing promises relevant speedups in some relevant cases.
Among the different financial applications that could benefit from the use of quantum
computing, we focus on the pricing of financial derivatives. We refer the reader to \cite{survey} for a recent survey of classical and quantum techniques
for option pricing.  \\ \\
 
 As it is well known in the literature (see for instance \cite{glasserman}), the pricing of financial derivatives can be formulated in terms of the computation of the expectation of the derivatives payoff with respect to a given probability measure. This computation can be very consuming in terms of computational resources and is typically performed by means of Monte Carlo (MC) methods. In the context of classical Monte Carlo (CMC) methods, the quantum computing community has proposed a quantum version which can obtain quadratic speedups for very general settings as indicated in \cite{quadratic_MC}. We will refer to such techniques as Quantum Accelerated Monte Carlo (QAMC). They have been successfully applied to problems of financial derivatives pricing (see \cite{Rebentrost_2018,Stamatopoulos_2020}). \\ \\
However, to the best of our knowledge, in the setups treated in the literature it is assumed that the price and the payoff of the derivative are strictly positive. If it is not the case, in order for the technique to work we need to divide the problem into two: an estimation of the derivatives price where the codomain is positive and a separate estimation of the derivatives price where the codomain is negative. Then, both solutions
are combined to obtain the final estimate. This article is centered around the idea of building an alternative protocol which avoids the hassle of separating into codomains.\\ \\

There are three main contributions in this article. First, the direct encoding protocol, which allows to load into the quantum circuit information about the sign of the expected payoff. Second, the modified real quantum amplitude estimation (mRQAE) which is an asymptotically optimal version of the real quantum amplitude estimation (RQAE) with the feature of recovering the sign of the underlying amplitude (see \cite{rqae}). Third, the whole pipeline, which exploits the synergies between both techniques.\\ \\

The manuscript is organised as follows. In Section \ref{sec:preliminaries} we begin by revising the main building components of CMC and QAMC. More precisely, Sections \ref{sec:CMC} and \ref{sec:quantum_pricing} contain a brief summary of the classical and quantum techniques to tackle the derivative pricing problem through Monte Carlo like techniques.
Then, Section \ref{sec:contrib} is devoted to the introduction of the new proposed strategy to perform a quantum Monte Carlo. The first part, in Section \ref{sec:new_encoding}, presents the direct encoding for negative payoffs along with its empirical evaluation. The second part, in Section \ref{sec:new_ae}, leverages the power of the mRQAE algorithm (see Appendix \ref{appendix:mRQAE}) for the pricing of derivative contracts with negative prices. Once again we perform an empirical assessment. In Section \ref{sec:conclusions} we summarize the main conclusions.

\section{Preliminaries}\label{sec:preliminaries}
In financial markets, a derivative is a financial contract whose value depends on the future performance of one or several assets usually referred to as underlying assets, or simply underlying (see, for example \cite{hull}). In this context, derivatives pricing consists of obtaining the price of the derivative at any previous date to maturity date. For this purpose, the uncertain future dynamics of the price of the underlying asset must be taken into account, which is usually modelled in terms of stochastic differential equations.\\

More precisely, if we denote by $S_t$ the price of the underlying asset at time $t$, a general It\^o process that satisfies the following Stochastic Differential Equation (SDE) can be considered:
\begin{equation} \label{eq:Ito_process}
    dS_t \, = \, \alpha(t,S_t) \, dt \, + \, \beta(t,S_t) \, dW_t, 
\end{equation}
where $\alpha$ and $\beta$ are real functions that represent the drift and the diffusion to be specified for the particular model, while $W_t$ denotes a Brownian motion in a particular probability space, so that $W_t$ follows a $\mathcal{N}(0,t)$ distribution (and its increment $dW_t$ follows a $\mathcal{N}(0,dt)$). \\

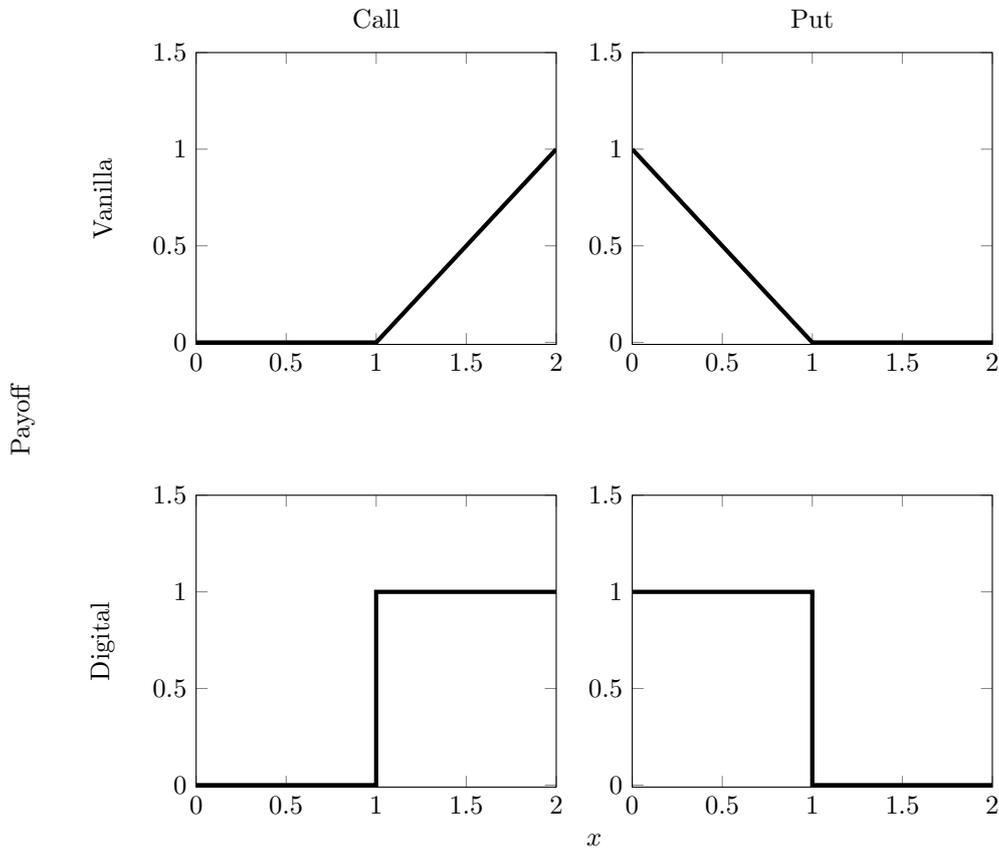
\begin{figure}[htbp!]
    \centering
\begin{tikzpicture}
\begin{groupplot}[group style={
group size=2 by 2,
vertical sep=2cm,horizontal sep=1cm},
width=0.4\textwidth,
xmin=0.,
xmax=2.,
ymin=-0.01,
ymax=1.5,
]
\nextgroupplot[title=Call,legend to name={groupplot1},ylabel=Vanilla,legend style={legend columns=3}]
\addplot+[no marks, ultra thick, color=black] coordinates {(0,0) (1,0) (2,1)};

\nextgroupplot[title=Put]
\addplot+[no marks, ultra thick, color=black] coordinates {(0,1) (1,0) (2,0)};
    
\nextgroupplot[ylabel=Digital]
\addplot+[const plot, no marks, ultra thick, color=black] coordinates {(0,0) (1,1) (2,1)};

\nextgroupplot
\addplot+[const plot, no marks, ultra thick, color=black] coordinates {(0,1) (1,0) (2,0)};

\end{groupplot}
\node[anchor=north] (title-x) at ($(group c1r2.south east)!0.5!(group c2r2.south west)-(0,0.5cm)$) {$x$};
\node[anchor=south, rotate=90] (title-y) at ($(group c1r1.south west)!0.5!(group c1r2.north west)-(2,0cm)$) {Payoff};

\path (group c1r2.north east) -- node[above=0.5cm]{\ref{groupplot1}} (group c2r2.north west);
\end{tikzpicture}
\caption{Payoff functions for digital and vanilla options with $K=1$.}\label{fig:payoffs}
\end{figure}

Another ingredient in the pricing of derivatives is the aforementioned payoff. We denote by $V_t$ the price of the derivative at time $t\in [0,T]$, where $T$ is the maturity date. Moreover, we assume the existence of a function $V$, such that $V_t=V(t,S_t)$, i.e., the value of the derivative depends on time and on the underlying asset price through a function $V$. Here we will focus on derivatives whose payoff only depends on the value of the asset at maturity, $V_T = F(S_T)$, where $F$ represents the payoff function. Next, if we denote the strike price by $K$, some examples of payoff functions are (see also Figure \ref{fig:payoffs}):

\begin{itemize}
    \item Vanilla call option: $F(x)= \max(x-K,0)$.
    \item Digital call option: $F(x)= 1_{x>K}$.
    \item Vanilla put option: $F(x)= \max(K-x,0)$.
    \item Digital put option: $F(x)= 1_{x<K}$.
\end{itemize}

By using mathematical finance tools, mainly martingale properties, It\^o's lemma and the Feynmann-Kàc theorem, the following expression for price of the derivative at time $t$ can be obtained (see, for example \cite{hull}): 
\begin{equation}\label{eq:BS-Feynmann}
    V_t=V(t,S_t) = \, e^{-r(T-t)}\mathbb{E}^Q[F(S_{T})|\mathcal{F}_t],
\end{equation}
where $\mathbb{E}^Q$ denotes the expectation under the risk neutral measure $Q$, $r$ is the constant risk-free interest rate, $F$ defines the payoff of the derivative and $\mathcal{F}_t$ denotes the $\sigma$-algebra containing the information until time $t$. In this way, expression (\ref{eq:BS-Feynmann}) indicates that the value of the derivative is the discounted price of the expected value of the payoff, conditioned to the current information of the market. \\

In view of the pricing expression (\ref{eq:BS-Feynmann}), the valuation of these financial derivatives mainly requires the computation of the involved expectation. Next, we briefly introduce one of the most popular techniques for computing such expectation, namely the Monte Carlo method. We revise the Classical Monte Carlo (CMC) in Section \ref{sec:CMC}  and the Quantum Accelerated Monte Carlo (QAMC) in Section \ref{sec:quantum_pricing}.
 \subsection{Classical Monte Carlo for derivatives pricing}\label{sec:CMC}
The CMC method for derivatives pricing in finance is composed of two steps:
\begin{enumerate}
    \item Simulation of sample paths of the underlying asset by means of the numerical solution of the SDE \eqref{eq:Ito_process}.
    \item Use of Monte Carlo integration to compute the expectation that appears in expression (\ref{eq:BS-Feynmann}).
\end{enumerate}
\subsubsection{Simulation of sample paths of the underlying price}\label{sec:classical_SDE}

For the simulation of the sample paths followed by the underlying asset price, there exist several numerical methods for solving the associated SDE \eqref{eq:Ito_process}. In order to illustrate the whole procedure we will use hereon the Euler-Maruyama method as an example. The application of the Euler-Maruyama scheme to the general SDE (\ref{eq:Ito_process}) leads to the expression \cite{pironneau}:
\begin{equation}\label{eq:euler_maruyama}
    S_{t_{j+1}} \; = \; S_{t_{j}}+\alpha(t_j,S_{t_j})  \Delta t \, + \, \beta(t_j, S_{t_j}) \sqrt{\Delta t} Z,
\end{equation}
where $Z$ is the standard normal random variable, i.e. with mean equal to $0$ and variance equal to $1$, and $\Delta t$ is the time step that is considered in the numerical method. Moreover, $\Delta t=(T-t)/M$, $M$ being the number of time steps, $T$ the maturity date, $t$ the initial time and $t_j = t + j\Delta t$ with $j=0,\ldots,M$. By using Equation \eqref{eq:euler_maruyama}, it is straightforward to produce samples of $S_T$, starting from a value of $S_t$ at time $t$. More precisely, we proceed as follows:
\begin{itemize}
    \item Start with the initial point $S_t$.
    \item Draw a sample of the standard normal random variable $Z$.
    \item Compute a sample of $S_{t+\Delta t}$ from the random sample generated in the previous step.
    \item Repeat the previous process, starting from the last calculated value, until a sample of $S_T$ is obtained.
\end{itemize}

\subsubsection{Integration by Monte Carlo}\label{sec:MC_integration}
If we repeat $N_C$ times the procedure described in Section \ref{sec:classical_SDE}, we obtain a set of $N_C$ paths of the price evolution and samples $S^i_T$, $i = 1,\ldots,N_C$ of the random variable $S_T$, which can be then used to estimate the expectation that appears in expression (\ref{eq:BS-Feynmann}) as follows:
\begin{equation}\label{eq:classical_MC}
    \mathbb{E}[F(S_T)|\mathcal{F}_t]= \dfrac{1}{N_C}\sum_{i = 1}^{N_C}F(S^i_T)+\epsilon_{\text{EM}}+\epsilon_{\text{CMC}}.
\end{equation}
In expresssion (\ref{eq:classical_MC}), $N_C$ is the number of samples $S^i_T$ generated by numerically solving the SDE, $F$ is the payoff function of the target derivatives contract, $\epsilon_{\text{CMC}}$ is the statistical error due to the Monte Carlo approximation of the expectation and $\epsilon_{\text{EM}}$ is the error comming from the discretization of the SDE, \emph{i.e.,} the error due to the Euler-Maruyama scheme. 
The statistical error $\epsilon_{\text{CMC}}$ scales as \cite{glasserman}:
\begin{equation}\label{eq:classical_MC_scaling}
    \epsilon_{\text{CMC}} \sim \dfrac{1}{\sqrt{N_{C}}}.
\end{equation}
The order of the error due to the Euler-Maruyama scheme is \cite{kloeden2013numerical}:
\begin{equation}
    \epsilon_{\text{EM}} \sim \Delta t \sim \dfrac{1}{M}.
\end{equation}
\subsection{Quantum Accelerated Monte Carlo for derivatives pricing}\label{sec:quantum_pricing}
The QAMC for pricing contains three main ingredients:
\begin{itemize}
    \item A quantum circuit which samples paths with the same probability as the classical circuit.
    \item An operator which encodes the payoff of the specific derivative contract into the quantum state.
    \item An amplitude estimation routine, which allows to retrieve the quantity of interest from the amplitude of a quantum state (and produces the actual speedup).
\end{itemize}
In Section \ref{sec:standard_encoding} we briefly discuss the first issue, while we reserve Section \ref{sec:ae} for the second and third issues.

\subsubsection{Quantum simulation}\label{sec:standard_encoding}
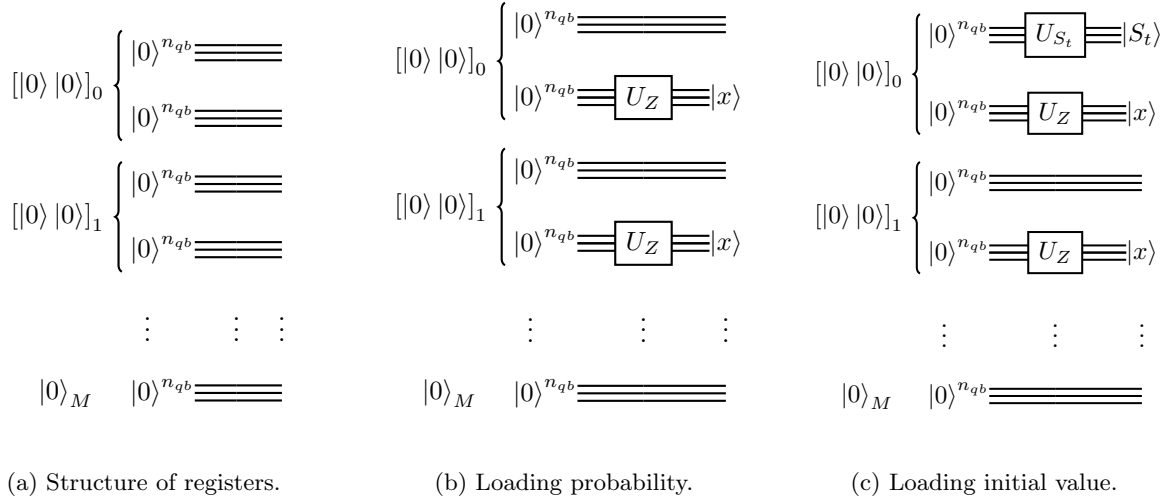
\begin{figure}[htbp!]
     \centering
     \begin{subfigure}[b]{0.3\textwidth}
         \centering
         \adjustbox{max width=0.99\textwidth}{\begin{quantikz}
    \lstick[wires=2]{$\left[\ket{0}\ket{0}\right]_0$}
    \push{\ket{0}^{n_{qb}}} &\qwbundle[alternate]{}&\qwbundle[alternate]{}\\
    \push{\ket{0}^{n_{qb}}} &\qwbundle[alternate]{}&\qwbundle[alternate]{}\\
    \lstick[wires=2]{$\left[\ket{0}\ket{0}\right]_1$}
    \push{\ket{0}^{n_{qb}}} &\qwbundle[alternate]{}&\qwbundle[alternate]{}\\
    \push{\ket{0}^{n_{qb}}} &\qwbundle[alternate]{}&\qwbundle[alternate]{}\\
    \lstick[wires=1]{\vdots}
    & \vdots & \vdots\\
    \lstick[wires=1]{$\ket{0}_{M}\quad$}
    \push{\ket{0}^{n_{qb}}} &\qwbundle[alternate]{}&\qwbundle[alternate]{}\\
 \end{quantikz}}
         \caption{Structure of registers.}
         \label{cir:structure}
     \end{subfigure}
     \hfill
     \begin{subfigure}[b]{0.3\textwidth}
         \centering
\adjustbox{max width=0.99\textwidth}{\begin{quantikz}
    \lstick[wires=2]{$\left[\ket{0}\ket{0}\right]_0$}
    \push{\ket{0}^{n_{qb}}} &\qwbundle[alternate]{}&\qwbundle[alternate]{}\\
    \push{\ket{0}^{n_{qb}}} & \gate{U_Z} \qwbundle[alternate]{} & \push{\ket{x}} \qwbundle[alternate]{}\\
    \lstick[wires=2]{$\left[\ket{0}\ket{0}\right]_1$}
    \push{\ket{0}^{n_{qb}}} &\qwbundle[alternate]{}&\qwbundle[alternate]{}\\
    \push{\ket{0}^{n_{qb}}} & \gate{U_Z} \qwbundle[alternate]{} & \push{\ket{x}} \qwbundle[alternate]{}\\
    \lstick[wires=1]{\vdots}
    & \vdots & \vdots\\
    \lstick[wires=1]{$\ket{0}_{M}\quad$}
    \push{\ket{0}^{n_{qb}}} &\qwbundle[alternate]{}&\qwbundle[alternate]{}\\
 \end{quantikz}}
         \caption{Loading probability.}
         \label{cir:loading_probability}
     \end{subfigure}
     \hfill
     \begin{subfigure}[b]{0.3\textwidth}
         \centering
\adjustbox{max width=0.99\textwidth}{\begin{quantikz}
    \lstick[wires=2]{$\left[\ket{0}\ket{0}\right]_0$}
    \push{\ket{0}^{n_{qb}}} & \gate{U_{S_t}} \qwbundle[alternate]{}&\push{\ket{S_t}} \qwbundle[alternate]{}\\
    \push{\ket{0}^{n_{qb}}} & \gate{U_Z} \qwbundle[alternate]{} & \push{\ket{x}} \qwbundle[alternate]{}\\
    \lstick[wires=2]{$\left[\ket{0}\ket{0}\right]_1$}
    \push{\ket{0}^{n_{qb}}} &\qwbundle[alternate]{}&\qwbundle[alternate]{}\\
    \push{\ket{0}^{n_{qb}}} & \gate{U_Z} \qwbundle[alternate]{} & \push{\ket{x}} \qwbundle[alternate]{}\\
    \lstick[wires=1]{\vdots}
    & \vdots & \vdots\\
    \lstick[wires=1]{$\ket{0}_{M}\quad$}
    \push{\ket{0}^{n_{qb}}} &\qwbundle[alternate]{}&\qwbundle[alternate]{}\\
 \end{quantikz}}
         \caption{Loading initial value.}
         \label{cir:loading_initial_value}
     \end{subfigure}
        \caption{Circuit initialisation.}
        \label{cir:initialisation}
\end{figure}

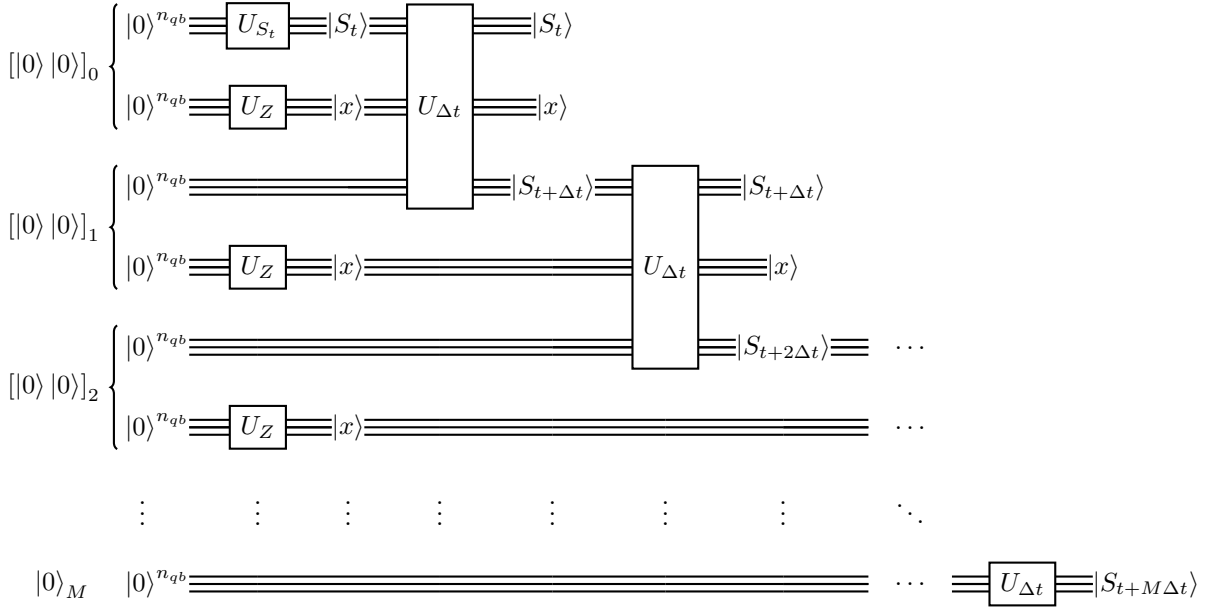
\begin{figure}[htbp!]
    \centering
\adjustbox{max width=\textwidth}{\begin{quantikz}
    \lstick[wires=2]{$\left[\ket{0}\ket{0}\right]_0$}
    \push{\ket{0}^{n_{qb}}} 
    & \gate{U_{S_t}}\qwbundle[alternate]{}     
    & \push{\ket{S_t}} \qwbundle[alternate]{}
    & \gate[wires=3]{U_{\Delta t}} \qwbundle[alternate]{}
    & \push{\ket{S_t}}\qwbundle[alternate]{}\\
    \push{\ket{0}^{n_{qb}}} 
    & \gate{U_Z} \qwbundle[alternate]{} 
    & \push{\ket{x}} \qwbundle[alternate]{}
    & \qwbundle[alternate]{}
    & \push{\ket{x}}\qwbundle[alternate]{}\\
    \lstick[wires=2]{$\left[\ket{0}\ket{0}\right]_1$}
    \push{\ket{0}^{n_{qb}}} 
    & \qwbundle[alternate]{}
    & \qwbundle[alternate]{}
    & \qwbundle[alternate]{}
    & \push{\ket{S_{t+\Delta t}}}\qwbundle[alternate]{}
    & \gate[wires=3]{U_{\Delta t}} \qwbundle[alternate]{}
    & \push{\ket{S_{t+\Delta t}}} \qwbundle[alternate]{}\\
    \push{\ket{0}^{n_{qb}}} 
    & \gate{U_Z} \qwbundle[alternate]{} 
    & \push{\ket{x}} \qwbundle[alternate]{}
    & \qwbundle[alternate]{}
    & \qwbundle[alternate]{}
    & \qwbundle[alternate]{}
    & \push{\ket{x}} \qwbundle[alternate]{}\\
    \lstick[wires=2]{$\left[\ket{0}\ket{0}\right]_2$}
    \push{\ket{0}^{n_{qb}}} 
    & \qwbundle[alternate]{}
    & \qwbundle[alternate]{}
    & \qwbundle[alternate]{}
    & \qwbundle[alternate]{}
    & \qwbundle[alternate]{}
    & \push{\ket{S_{t+2\Delta t}}} \qwbundle[alternate]{}
    & \push{\quad\hdots\quad} \qwbundle[alternate]{}\\
    \push{\ket{0}^{n_{qb}}} 
    & \gate{U_Z} \qwbundle[alternate]{} 
    & \push{\ket{x}} \qwbundle[alternate]{}
    & \qwbundle[alternate]{}
    & \qwbundle[alternate]{}
    & \qwbundle[alternate]{}
    & \qwbundle[alternate]{}
    & \push{\quad\hdots\quad} \qwbundle[alternate]{}\\
    \lstick[wires=1]{\vdots}
    & \vdots 
    & \vdots
    & \vdots 
    & \vdots 
    & \vdots 
    & \vdots
    & \ddots\\
    \lstick[wires=1]{$\ket{0}_{M}\quad$}
    \push{\ket{0}^{n_{qb}}}
    & \qwbundle[alternate]{}
    & \qwbundle[alternate]{}
    & \qwbundle[alternate]{}
    & \qwbundle[alternate]{}
    & \qwbundle[alternate]{}
    & \qwbundle[alternate]{}
    & \push{\quad\hdots\quad} \qwbundle[alternate]{}
    & \gate[wires=1]{U_{\Delta t}} \qwbundle[alternate]{}
    & \push{\ket{S_{t+M\Delta t}}} \qwbundle[alternate]{}\\
 \end{quantikz}}
    \caption{Sketch description of the construction of the oracle defined in Equation \eqref{eq:path_superposition}.}
    \label{reg:all_steps}
\end{figure}

The QAMC algorithm begins by creating a state in superposition where the probabilities of each path match those of the classical process discretized  by using some numerical scheme such as the Euler-Maruyama formula. Alternative methods for the numerical solution of SDEs  with different orders of convergence can be considered (see, for example \cite{kloeden2013numerical}). In order to build the algorithm, $M+1$ different registers are needed, one per time step. The first $M$ registers are composed of two registers of $n_{qb}$ qubits each (see Figure \ref{cir:structure}):
\begin{equation}\label{eq:original_register}
    \left[\ket{0}\ket{0}\right]_0\otimes\left[\ket{0}\ket{0}\right]_1\otimes\cdots\otimes\ket{0}_{M},
\end{equation}
where $\left[\ket{0}\ket{0}\right]_m = \left[\ket{0}^{\otimes n_{qb}}\ket{0}^{\otimes n_{qb}}\right]_m$, $m = 0,\ldots,M$ and  $\ket{0}_{M} = \ket{0}^{\otimes n_{qb}}$.\\ \\
Each of the individual registers $\ket{0}^{\otimes n_{qb}}$ will be used to represent a decimal number. For simplicity, it can be understood as a single precision register. In order to generate a state in superposition which matches the probabilities of each path defined by the Equation \eqref{eq:euler_maruyama}, we need a standard normal sample generator. In the QAMC algorithm, this generator is represented by the unitary operator $U_{Z}$ which performs the following transformation:
\begin{equation} \label{transform}
    U_{Z}\ket{0}\ket{0} = \sum_{j = 1}^{J}\sqrt{p_{Z}(x_j)}\ket{0}\ket{x_j} = \ket{0}\ket{x},
\end{equation}
where $x$ is a set of $J$ numbers that can be represented by the $n_{qb}$ qubits from the individual registers and $p_{Z}(x)$ is a discretized version of the standard normal probability distribution defined in the set of points $x= \{x_1,x_2,...,x_{J}\}$. Note that, in general, $J$ does not need to be equal to $2^{n_{qb}}$. The efficiency of the transformation \eqref{transform} is crucial for the overall efficiency of the algorithm. In the best case, this efficiency can be achieved using $O(\log_2 (J))$ gates (see \cite{prob_loading}). In the worst case, it can be achieved in $O(J\log_2(J))$ combining the results in \cite{prob_loading} and \cite{Shende_2006}.\\ \\
The first step requires applying the operator $U_Z$ to one of the members of all pairs $\left[\ket{0}\ket{0}\right]_i$, thus obtaining the state:
\begin{eqnarray}
        & & \left[U_Z\ket{0}\ket{0}\right]_0\otimes\left[U_Z\ket{0}\ket{0}\right]_1\otimes\cdots\otimes\ket{0}_{M} \nonumber\\ & = & \left[\sum_{j = 1}^{J}\sqrt{p_Z(x_j)}\ket{0}\ket{x_j}\right]_0\otimes\left[\sum_{j = 1}^{J}\sqrt{p_Z(x_j)}\ket{0}\ket{x_j}\right]_1\otimes\cdots\otimes\ket{0}_{M}.\nonumber
\end{eqnarray}
In this configuration, the amplitudes encode the square root of the probabilities for all the different combinations of $x$ in the different steps. Next, the left register in the pair $\left[\ket{0}\ket{0}\right]_0$ has to be initialised to $\ket{S_t}$:
\begin{equation}\label{eq:initialization_S_t}
    \left[U_{S_t}\ket{0}\ket{y}\right]_0 =  \left[\ket{S_t}\ket{y}\right]_0,\quad \forall \; \ket{y}.
\end{equation}
Figure \ref{cir:initialisation} depicts schematically this process.
Once the circuit is correctly initialised, an evolution operator must be applied. This evolution operator $U_{\Delta t}$ acts upon three individual registers as follows:
\begin{equation}\label{eq:update_rule}
    U_{\Delta t}\left[\ket{S_{t+m\Delta t}}\ket{x}\right]_m\left[\ket{0}\right]_{m+1}\longrightarrow\left[\ket{S_{t+m\Delta t}}\ket{x}\right]_m\left[\ket{S_{t+(m+1)\Delta t}(S_{t+m\Delta t},x)}\right]_{m+1},
\end{equation}
where the update rule is given, for instance, by Equation \eqref{eq:euler_maruyama}. Note that other update rules can be used instead. After repeatedly applying the operator $U_{\Delta t}$ defined in Equation \eqref{eq:update_rule}, the final quantum state $\ket{S}$ :
\begin{equation}\label{eq:path_superposition}
    \ket{S} := U_S\ket{0} = \sum_{k = 0}^{\mathcal{K}-1}\sqrt{p_S(S_k)}\ket{S_k},
\end{equation}
where $M$ is the number of time steps, $\mathcal{K} = J^M$ are the number of possible paths defined by the given (space and time) discretization and $p_S(S_k)$ is the probability of generating the path $S_k$.
Figure \ref{reg:all_steps} depicts schematically this process.
So far, a quantum circuit has been built which samples paths with the same probability as the classical circuit does. Moreover, the computational cost of one execution of the circuit is equivalent to one execution of the classical circuit, \emph{i.e.}, the number of gates needed to sample one path from the classical and the quantum circuit is ``the same'', since the classical circuit can always be translated to a quantum one using Toffoli gates (see \cite{nielsen_chuang}, for example). However, note that classical and quantum gates are not directly comparable.

\subsubsection{Amplitude estimation}\label{sec:ae}
As discussed in the previous section, up to this point the quantum and the classical circuit have the same complexity. Nevertheless, when the error correction is taken into consideration, the current quantum gates are much slower than the analogous classical ones. Next, the mechanism that produces an speedup is briefly detailed.\\ \\
 Once the state $\ket{S}$ in Equation \eqref{eq:path_superposition} is generated, the next step is to define the operator $U_{\sqrt{F}}$ such that pushes the square root of the derivatives payoff $F$ into the amplitude. For this reason, we will call this way of encoding \emph{square root encoding}. For this purpose, an additional single qubit register is needed:
\begin{equation}\label{eq:final_state}
	\begin{aligned}
    \ket{\sqrt{F}} = U_{\sqrt{F}}\ket{S}\ket{0} = &\dfrac{1}{\|\sqrt{F(S)}\|_{\infty}}\sum_{k= 0}^{\mathcal{K}-1}\sqrt{p_S(S_k)F(S_k)}\ket{S_k}\ket{0}\\
    &+\sqrt{p_S(S_k)\left(1-F(S_k)\right)}\ket{S_k}\ket{1}.
\end{aligned}
\end{equation}

\begin{figure}
 	\centering
          \begin{quantikz}
 	\lstick[wires=1]{$\ket{0}$}  
 	&\qwbundle[alternate]{}
 	&\gate[wires=1]{U_{S}}\qwbundle[alternate]{}
 	&\gate[wires=2]{U_{\sqrt{F}}}\qwbundle[alternate]{}
 	&\qwbundle[alternate]{}
 	\rstick[wires=3]{$\begin{aligned}\dfrac{1}{\|\sqrt{F(S)}\|_{\infty}}\sum_{k= 0}^{\mathcal{K}-1}&\sqrt{p_S(S_k)F(S_k)}\ket{S_k}\ket{0}\\+&\sqrt{p_S(S_k)\left(1-F(S_k)\right)}\ket{S_k}\ket{1}\end{aligned}$}\\
 	\lstick[wires=1]{$\ket{0}$}
 	&\qw
 	&\qw
 	&\qw
 	&\qw\\
 \end{quantikz}
    \caption{Scheme of the generation of the oracle in the square root encoding. The gate $U_S$ corresponds to Equation \eqref{eq:path_superposition}. The gate $U_{\sqrt{F}}$ corresponds to Equation \eqref{eq:final_state}.}\label{fig:final_state}
\end{figure}
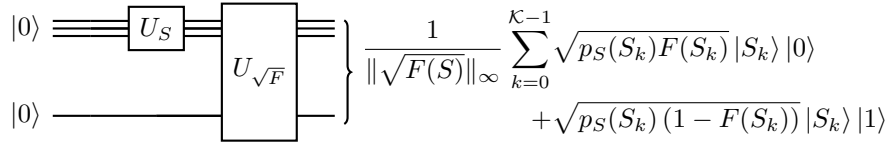

 Moreover, it is tacitly assumed that the operator $U_{\sqrt{F}}$ can be efficiently implemented. Figure \ref{fig:final_state} depicts schematically the overall process. We will denote by $U_{\text{SRE}}$ the combination of the path generating oracle $U_S$ and the payoff oracle $U_{\sqrt{F}}$:
 \begin{equation}\label{eq:sqrt_encoding}
     U_{\text{SRE}} := U_{\sqrt{F}}U_S.
 \end{equation}

The probability of measuring zero in the rightmost register after the application of $ U_{\text{SRE}}$ is given by:
	\begin{equation}
		P_{\ket{0}} = \dfrac{1}{\|\sqrt{F(S)}\|_{\infty}^2}\sum_{k = 0}^{\mathcal{K}-1}\left|p_S(S_k)F(S_k)\right|.
	\end{equation}

Hence, getting an estimation $\widetilde{P}_{\ket{0}}$ of $P_{\ket{0}}$ yields an estimation of the expectation in Equation \eqref{eq:BS-Feynmann} except for the normalization constants:
\begin{equation}\label{eq:quantum_estimation}
    \mathbb{E}[F(S_T)| \mathcal{F}_t] = \sum_{k = 0}^{\mathcal{K}-1}p_S(S_k)F(S_k)+\epsilon_{\mathcal{K}}+\epsilon_{\text{EM}} \approx \|\sqrt{F(S)}\|_{\infty}^2\widetilde{P}_{\ket{0}}+\epsilon_{\mathcal{K}}+\epsilon_{\text{EM}}+\epsilon_{\text{QAMC}},
\end{equation}
where $\sum_{k = 0}^{\mathcal{K}-1}p_S(S_k)F(S_k)$ is the discretized expectation, $\epsilon_{\text{EM}}$ is same Euler-Maruyama error as in Equation \eqref{eq:classical_MC}, $\epsilon_{\mathcal{K}}$ is the discretization error, $ \widetilde{P}_{\ket{0}}$ is an estimation of the probability of obtaining zero in the last register
and $\epsilon_{\text{QAMC}}$ is the sampling error given by,
\begin{equation}\label{eq:epsilon_square}
	\epsilon_{\text{QAMC}} = \left|\|\sum_{k = 0}^{\mathcal{K}-1}p_S(S_k)F(S_k) - \sqrt{F(S)}\|_{\infty}^2\widetilde{P}_{\ket{0}}\right|.
\end{equation}
The discretization error $\epsilon_{\mathcal{K}}$ is intimately related with the discretization of the standard normal probability distribution. When we use a fine-grain discretization this error can be considered negligible compared with the other ones. This is the case of the CMC, where we typically use $32$ bits (single precision) to represent the standard normal. For this reason and to avoid confusion we have omitted any reference to the discretization error in Equation \eqref{eq:classical_MC}.\\

By using amplitude estimation techniques, we know that the sampling error $\epsilon_{\text{QAMC}}$ is of order \cite{Brassard_2002}:
\begin{equation}\label{eq:error_estimation}
    \epsilon_{\text{QAMC}} \sim \dfrac{1}{N_Q},
\end{equation}
with $N_Q$ being the number of calls to the oracle defined by Equation \eqref{eq:sqrt_encoding}. Recall that this oracle is strictly the same as in the classical algorithm described in Section \ref{sec:classical_SDE}. Thus, each call to the oracle $U_{\text{SRE}}$ is equivalent to $M$ steps of the Euler-Maruyama scheme. Technically speaking, since the application of amplitude estimation techniques requires the use of the adjoint of the oracle $U_{\text{SRE}}$ we are assuming that both have the same cost. \\ \\
In Table \ref{tab:quantum_error_comparison} we show the computational cost of CMC and QAMC, measured in terms to the number of queries. It can be easily seen that the QAMC performs quadratically better than the CMC. Moreover, the same scaling applies when we increase the number of dimensions.
\begin{table}[htbp!]
    \centering
    \begin{tabular}{|c|c|}
    \hline
         &  Error \\
         \hline
        CMC &$O(1/\sqrt{N_C})$\\
        \hline
        QAMC &$O(1/N_Q)$\\
        \hline
    \end{tabular}
    \caption{Comparison of the order of the errors for the CMC and the QAMC.}
    \label{tab:quantum_error_comparison}
\end{table}

\subsubsection{QAMC simplifications and practical implementation}\label{sec:simplifications}
\begin{figure}[htbp!]
    \centering
\begin{tikzpicture}
\begin{groupplot}[group style={
group size=2 by 2,
vertical sep=2cm,horizontal sep=1cm},
width=0.4\textwidth,
xmode = log,
ymode = log,
xmin=10e-8,
xmax=0.01,
ymin=10e2,
ymax=5*10e6,
x dir=reverse,
]
\nextgroupplot[title=Call,legend to name={groupplot2},ylabel=Vanilla,legend style={legend columns=2}]
    \addplot [color=green,mark=*] 
  plot [error bars/.cd, x dir=both, x explicit, y dir=both, y explicit]
  table [x=error_median,y=oracle_calls_median,x error plus=error_75, x error minus=error_25, y error plus=oracle_calls_75, y error minus=oracle_calls_25]
  {data/mIQAE-Square-European_Call_Option-100.dat};
    \addplot[color=gray,domain=5*10e-2:10e-6,style=dashed]{1/(sqrt(x)};
\addlegendentry{mIQAE}

\nextgroupplot[title=Put]
    \addplot [color=green,mark=*] 
  plot [error bars/.cd, x dir=both, x explicit, y dir=both, y explicit]
  table [x=error_median,y=oracle_calls_median,x error plus=error_75, x error minus=error_25, y error plus=oracle_calls_75, y error minus=oracle_calls_25]
  {data/mIQAE-Square-European_Put_Option-100.dat};
    \addplot[color=gray,domain=5*10e-2:10e-6,style=dashed]{1/(sqrt(x)};
    
\nextgroupplot[ylabel=Digital]
    \addplot [color=green,mark=*] 
  plot [error bars/.cd, x dir=both, x explicit, y dir=both, y explicit]
  table [x=error_median,y=oracle_calls_median,x error plus=error_75, x error minus=error_25, y error plus=oracle_calls_75, y error minus=oracle_calls_25] {data/mIQAE-Square-Digital_Call_Option-100.dat};
    \addplot[color=gray,domain=5*10e-2:10e-6,style=dashed]{1/(sqrt(x)};
    
\nextgroupplot
    \addplot [color=green,mark=*] 
  plot [error bars/.cd, x dir=both, x explicit, y dir=both, y explicit]
  table [x=error_median,y=oracle_calls_median,x error plus=error_75, x error minus=error_25, y error plus=oracle_calls_75, y error minus=oracle_calls_25] {data/mIQAE-Square-Digital_Put_Option-100.dat};
\end{groupplot}
\node[anchor=north] (title-x) at ($(group c1r2.south east)!0.5!(group c2r2.south west)-(0,0.5cm)$) {$\epsilon_{\text{QAMC}}$};
\node[anchor=south, rotate=90] (title-y) at ($(group c1r1.south west)!0.5!(group c1r2.north west)-(2,0cm)$) {$N_Q$};

\path (group c1r2.north east) -- node[above=0.5cm]{\ref{groupplot2}} (group c2r2.north west);
\end{tikzpicture}
\caption{Absolute error between the QAMC algorithm and the discretized expectation versus the respective number of calls to the oracle for different precisions $\epsilon$. The dots represent the medians and the error bars the $25$ and $75$ percentiles. Each of the panels corresponds to the payoff of different options. The experiments have been performed using the square root encoding. }\label{fig:probability_loading_true}
\end{figure}
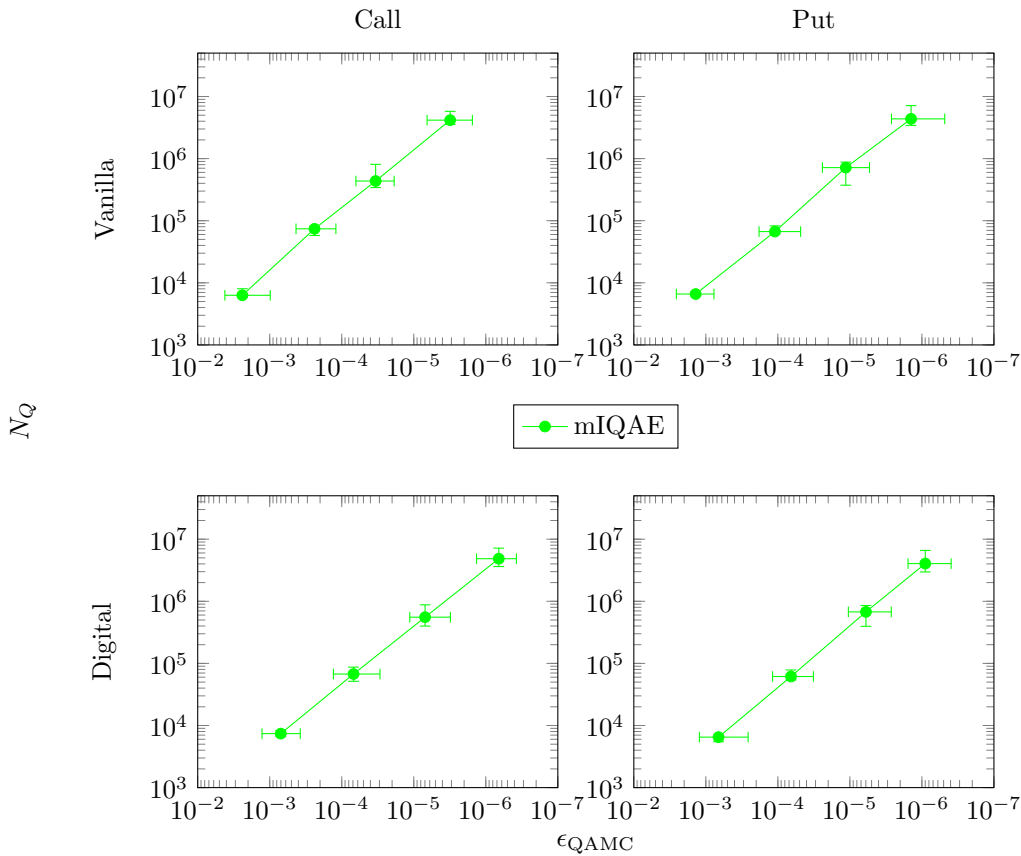

So far, we have described the general setup of QAMC for pricing. A rough estimation indicates that we would require the order of hundreds or thousands of logical qubits to build the algorithm with single precision registers and a few time steps. With the current hardware, this is not feasible (see \cite{list_quantum_processors}). Hence, in order to conceptually test this technique, we need to perform several simplifications.\\ \\
If we assume only European payoffs we can make the first simplification since we do not need to store the whole paths for the underlying. Instead, we will consider that we have just one register which encodes the value of the underlying and we will rewrite it on each step of the Euler-Maruyama scheme.\\
The next simplification would be reducing as much as possible the number of time steps. In the limit, we could perform a single time step. Nevertheless, in numerical schemes such as Euler-Maruyama, a very big time step produces a very big error. For that reason, we restrict ourselves to models where we know how to do exact simulation, thus avoiding the need of doing several steps. This happens when we can obtain the exact solution of the governing SDE. Thus, in this work, we will consider the classical (and well known) model given by the following Black-Scholes SDE under the risk neutral measure \cite{BS_first} 
\begin{equation}\label{eq:black_scholes_SDE}
    dS_t \; = \; r S_t \, dt \, + \, \sigma S_t 
    \, dW_t,
\end{equation}
where $r$ denotes the risk-free rate and $\sigma$ is the volatility of the underlying asset price. As the expression of the exact solution of SDE \eqref{eq:black_scholes_SDE} is known,
starting for a given value $S_t$, the simulation of the random variable $S_T$ under the risk-neutral measure can be exactly carried out in one step by:
\begin{equation}\label{eq:BS_exact_simulation}
    S_T = S_t \exp\left(\left(r-\dfrac{1}{2}\sigma^2\right)(T-t)+\sigma Z \sqrt{T-t}\right).
\end{equation}
Under the previously simplified setting, we will only need three (or even two) registers. The first one for storing the initial value of the underlying $S_t$, the second one would be the register for the standard normal $Z$ and the third one for the final value $S_T$. Since we only perform one step, actually storing $S_t$ is not strictly necessary as it can be hardcoded in the operator $U_{\Delta t}$ leaving us with only two registers.\\ \\
In yet another simplification, we assume that, instead of having a unitary operator $U_Z$ which encodes the standard normal, we have an analogous unitary $U_{\text{BS}}$ which encodes the Black-Scholes distribution $p_{\text{BS}}$. With this last simplification, we only need a single register to perform the whole simulation. Note that, regardless all the simplifications, the algorithm is conceptually the same: we have a quantum circuit specified by the oracle $U_S = U_{\text{BS}}$ which generates samples for the underlying price at maturity, $S_T$, with the correct probability distribution.\\ \\
Throughout the manuscript, we will show different numerical experiments, all of them performed under the simplifications described in this section. For the theoretical discussions, we will continue referring to the general setting from previous sections.\\ \\
To wrap up this section, in Figure \ref{fig:probability_loading_true} we show the results of the QAMC for different payoffs when the modified iterative amplitude estimation (mIQAE) algorithm (see \cite{Fukuzawa_2023}) is used. The mIQAE is considered the current state of the art with regard to amplitude estimation algorithms. For all the experiments we have encoded the Black-Scholes probability distribution with risk-free rate $0.01$ and volatility $0.5$. Moreover, we have considered a one year maturity and an initial underlying value of $1.0$. For the discretization of the distribution we have considered $32$ points between $0.01$ and $5.0$ which requires the use of $5$ qubits. 

\section{Alternative schedule for QAMC}\label{sec:contrib} 
As it is shown in Section \ref{sec:ae}, by sampling from the quantum circuit we can obtain an estimation $\widetilde{P}_{\ket{0}}$ of:
\begin{equation}
   \dfrac{1}{\|\sqrt{F(S)}\|_{\infty}}\sum_{k = 0}^{\mathcal{K}-1}\left|p_S(S_k)F(S_k)\right|.
\end{equation}

\begin{figure}[htbp!]
\centering
\begin{tikzpicture}
\begin{axis}[
xmin=0.5,
xmax=2.5,
ymin=-1,
ymax=1.,
xlabel=$x$,
ylabel=Payoff,
]
\addplot+[no marks, ultra thick, color=black] coordinates {(0.5,-1) (1.5,0) (2.5,1)};
\end{axis}
\end{tikzpicture}
\caption{Linear payoff with $K = 1.5$.}\label{fig:linear_payoff}
\end{figure}
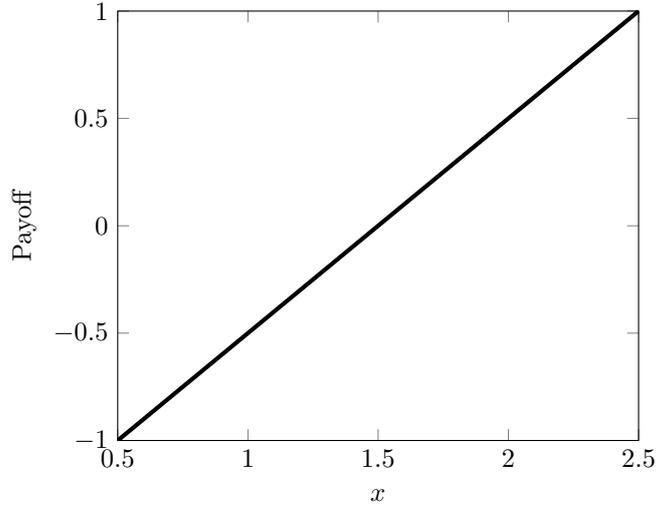

Nevertheless, it is important to note that, for derivatives whose payoffs can become negative, the naive use of this method will not yield to correct prices approximations. In order to illustrate this, suppose that there is a payoff of the form (see Figure \ref{fig:linear_payoff}):
\begin{equation}
    F(S_T) = S_T-K,
\end{equation}
with $T$ being the maturity of the contract, $S_T$ the price of the underlying at maturity and $K$ the strike price of the contract (see \cite{survey} for details).

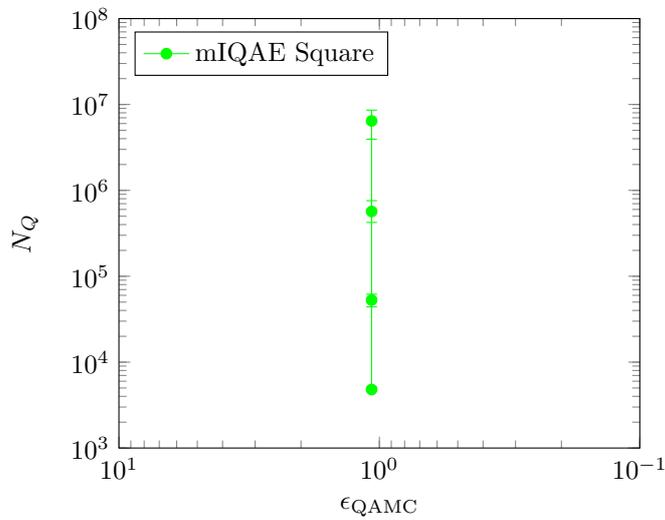
\begin{figure}[htbp!]
\centering
\begin{tikzpicture}
\begin{axis}[
xmode = log,
ymode = log,
ymin=10e2,
ymax=10e7,
xmin=10e-2,
xmax=10,
max space between ticks=50pt,
x dir=reverse,
xlabel=$\epsilon_{\text{QAMC}}$,
ylabel=$N_Q$,
legend pos=north west,
]\addplot [color=green,mark=*] 
  plot [error bars/.cd, x dir=both, x explicit, y dir=both, y explicit]
  table [x=error_median,y=oracle_calls_median,x error plus=error_75, x error minus=error_25, y error plus=oracle_calls_75, y error minus=oracle_calls_25] {data/mIQAE-Square-Futures-100.dat};
  \addlegendentry{mIQAE Square}
\end{axis}
\end{tikzpicture}
\caption{Absolute error between the QAMC algorithm and the discretized expectation of an option with a payoff $(S_T-K)$ with $K = 1.5 S_t$ versus the number of calls to the oracle for different values of precision $\epsilon$. The dots represent the medians and the error bars the $25$ and $75$ percentiles. The experiments have been performed using the square root encoding. }\label{fig:sum_absolute_values}
\end{figure}

Figure \ref{fig:sum_absolute_values} shows the results for the the square root encoding combined with the mIQAE for a naive implementation of QAMC. It illustrates that there is no convergence to the correct value because of the presence of the absolute value.\\

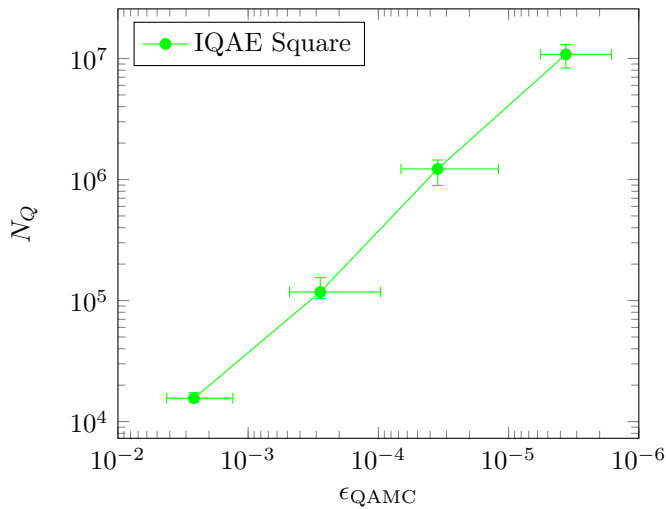
\begin{figure}[htbp!]
\centering
\begin{tikzpicture}
\begin{axis}[
xmode = log,
ymode = log,
xmin=10e-7,
xmax=0.01,
max space between ticks=50pt,
x dir=reverse,
xlabel=$\epsilon_{\text{QAMC}}$,
ylabel=$N_Q$,
legend pos=north west,
]\addplot [color=green,mark=*] 
  plot [error bars/.cd, x dir=both, x explicit, y dir=both, y explicit]
  table [x=error_median,y=oracle_calls_median,x error plus=error_75, x error minus=error_25, y error plus=oracle_calls_75, y error minus=oracle_calls_25] {data/mIQAE-Square-Parts-Futures-100.dat};
  \addlegendentry{IQAE Square}
\end{axis}
\end{tikzpicture}
\caption{Absolute error between the QAMC algorithm and the discretized expectation of an option with a payoff $(S_T-K)$ with $K = 1.5 S_t$ versus the number of calls to the oracle for different values of precision $\epsilon$. The dots represent the medians and the error bars the $25$ and $75$ percentiles. The experiments have been performed using the square root encoding separating the positive and negative parts of the payoff. }\label{fig:future_sqrt_encoding_separate}
\end{figure}

In order to avoid the errors introduced by the presence of the absolute values in the QAMC, whenever we have a payoff that is potentially negative we must divide our problem into two distinct problems. On the one hand, we must define the positive part of our target function:
\begin{equation}
    F_+(S_T) = \max(F(S_T),0).
\end{equation}
On the other hand, we define the negative part of our target function:
\begin{equation}
    F_-(S_T) = \abs{\min(F(S_T),0)}.
\end{equation}
Therefore, we can express our payoff as a linear combination of the positive and negative parts:
\begin{equation}
    F(S_T) = F_+(S_T) - F_-(S_T).
\end{equation}
In terms of estimation we need to perform a separate estimation of both the positive and negative parts and then combine the results. Applying this decomposition, the mIQAE provides the results in Figure \ref{fig:future_sqrt_encoding_separate}.

In Sections \ref{sec:new_encoding} and \ref{sec:new_ae} we develop a new strategy which does not require the user to separate the problem into two. On the one hand, a new encoding is proposed. On the other hand, a different amplitude estimation technique is employed.

\subsection{Direct encoding}\label{sec:new_encoding}
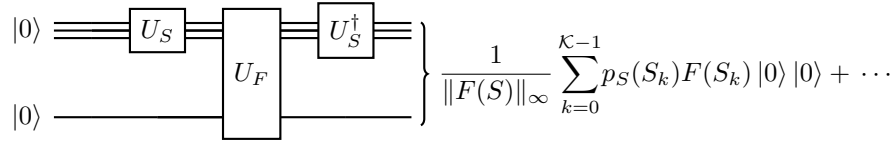
\begin{figure}[htbp!]
    \centering
         \begin{quantikz}
    \lstick[wires=1]{$\ket{0}$}  
    &\qwbundle[alternate]{}
    &\gate[wires=1]{U_{S}}\qwbundle[alternate]{}
    &\gate[wires=2]{U_{F}}\qwbundle[alternate]{}
    &\gate[wires=1]{U^\dagger_{S}}\qwbundle[alternate]{}
    &\qwbundle[alternate]{}
    \rstick[wires=3]{$\begin{aligned}\dfrac{1}{\|F(S)\|_{\infty}}\sum_{k = 0}^{\mathcal{K}-1}&p_S(S_k)F(S_k)\ket{0}\ket{0}+\,\cdots \end{aligned}$}\\
    \lstick[wires=1]{$\ket{0}$}
    &\qw
    &\qw
    &\qw
    &\qw
    &\qw\\
 \end{quantikz}
     \caption{Scheme of the generation of the oracle in the direct encoding. The gate $U_S$ corresponds to Equation \eqref{eq:path_superposition}. The gate $U_{F}$ corresponds to Equation \eqref{eq:final_state2}.}
    \label{fig:final_state_protocol_1}
\end{figure}

The \emph{direct encoding} algorithm starts from the same initial state $\ket{S}$:
\begin{equation}\label{eq:path_superposition2}
    \ket{S} = U_S\ket{0} = \sum_{k = 0}^{\mathcal{K}-1}\sqrt{p_S(S_k)}\ket{S_k},
\end{equation}
where $\mathcal{K}$ is again the number of possible paths defined by the given discretization. The next step is to define the operator $U_{F}$ such that pushes the payoff without squared roots into the amplitude. For this purpose, an additional single qubit register is needed, so that:
\begin{equation}\label{eq:final_state2}
	\begin{aligned}
    \ket{F} = U_{F}\ket{S}\ket{0} = \dfrac{1}{\|F(S)\|_{\infty}}\sum_{k = 0}^{\mathcal{K}-1}&\sqrt{p_S(S_k)}F(S_k)\ket{S_k}\ket{0}\\+&\sqrt{p_S(S_k)}\left(1-F(S_k)\right)\ket{S_k}\ket{1}.
    \end{aligned}
\end{equation}
Next, we apply the inverse of the $U_S$ unitary on the state $\ket{F}$ (see Figure \ref{fig:final_state_protocol_1}), thus getting:
\begin{equation}\label{eq:projection}
    U_S^\dagger\ket{F} = \dfrac{1}{\|F(S)\|_{\infty}} \sum_{k = 0}^{\mathcal{K}-1}p_S(S_k)F(S_k) \ket{0}\ket{0} +\, \cdots \,.
\end{equation}
We will denote by $U_{\text{DE}}$ the application of the whole pipeline:
 \begin{equation}\label{eq:direct_encoding}
     U_{\text{DE}} := U_S^\dagger U_{F}U_S.
 \end{equation}
The square root probability of measuring the eigenstate zero is:
\begin{equation}\label{eq:projection_zero}
    \sqrt{P_{\overline{\ket{0}}}} = \left|\bra{0}U_{\text{DE}}\ket{0}\right| = \dfrac{1}{\|F(S)\|_{\infty}}\left| \sum_{k = 0}^{\mathcal{K}-1}p_S(S_k)F(S_k)\right|.
\end{equation}
Note the difference between $P_{\ket{0}}$ from the square root encoding in Equation \eqref{eq:quantum_estimation} and $P_{\overline{\ket{0}}}$ from the direct encoding in Equation \eqref{eq:projection_zero}. The former one refers to the probability of measuring zero in the last register when the unitary $U_{\text{SRE}}$ is applied. The latter refers to the probability of measuring the eigenstate zero when the unitary $U_{\text{DE}}$ is applied.
Finally, we apply an amplitude estimation algorithm to state zero of the oracle $U_{\text{DE}}$ to obtain an estimate $\widetilde{P}_{\overline{\ket{0}}}$ of $P_{\overline{\ket{0}}}$, thus
getting an estimation of the expectation in Equation \eqref{eq:BS-Feynmann}:
\begin{equation}\label{eq:quantum_estimation2}
\mathbb{E}[F(S_T)| \mathcal{F}_t] = \sum_{k = 0}^{\mathcal{K}-1}p_S(S_k)F(S_k)+\epsilon_{\mathcal{K}}+\epsilon_{\text{EM}} \approx \|{F(S)}\|_{\infty}\sqrt{\widetilde{P}_{\overline{\ket{0}}}}+\epsilon_{\mathcal{K}}+\epsilon_{\text{EM}}+\epsilon_{\text{QAMC}},
\end{equation}
where $\sum_{k = 0}^{\mathcal{K}-1}p_S(S_k)F(S_k)$ is the discretized expectation, $\epsilon_{\text{EM}}$ is same Euler-Maruyama error as in Equation \eqref{eq:classical_MC}, $\epsilon_{\mathcal{K}}$ is the discretization error, $ \widetilde{P}_{\ket{0}}$ is an estimation of the probability of obtaining zero in the last register and $\epsilon_{\text{QAMC}}$ is the sampling error given by,
	\begin{equation}\label{eq:epsilon_direct}
		\epsilon_{\text{QAMC}} = \left|\sum_{k = 0}^{\mathcal{K}-1}p_S(S_k)F(S_k)-\|{F(S)}\|_{\infty}\sqrt{\widetilde{P}_{\overline{\ket{0}}}}\right|.
\end{equation}

 This procedure allows pricing options with negative payoffs when the expected value is positive. However, the presence of the outer absolute value in Equation \eqref{eq:projection_zero} still prevents from a correct estimation when negative expectations arise.

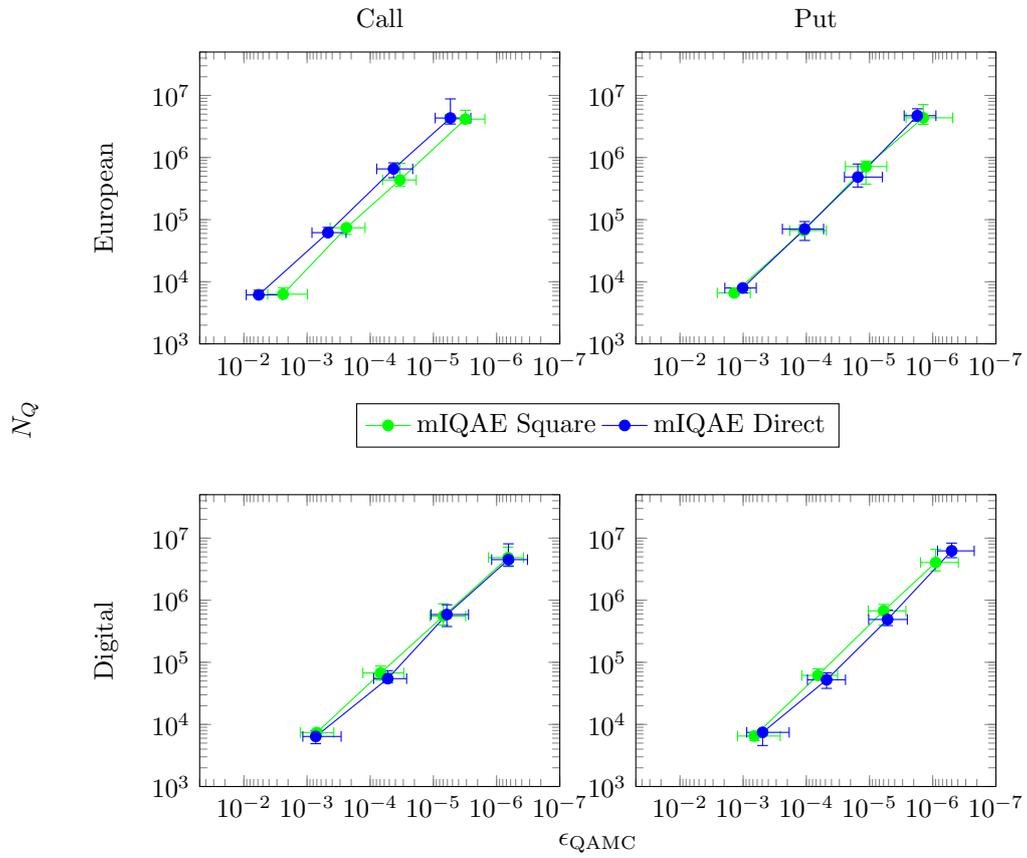
\begin{figure}[htbp!]
    \centering
\begin{tikzpicture}
\begin{groupplot}[group style={
group size=2 by 2,
vertical sep=2cm,horizontal sep=1cm},
width=0.4\textwidth,
xmode = log,
ymode = log,
xmin=10e-8,
xmax=0.05,
ymin=10e2,
ymax=5*10e6,
x dir=reverse,
]
\nextgroupplot[title=Call,legend to name={groupplot3},ylabel=European,legend style={legend columns=3}]
    \addplot [color=green,mark=*] 
  plot [error bars/.cd, x dir=both, x explicit, y dir=both, y explicit]
  table [x=error_median,y=oracle_calls_median,x error plus=error_75, x error minus=error_25, y error plus=oracle_calls_75, y error minus=oracle_calls_25]
  {data/mIQAE-Square-European_Call_Option-100.dat};
      \addplot [color=blue,mark=*] 
  plot [error bars/.cd, x dir=both, x explicit, y dir=both, y explicit]
  table [x=error_median,y=oracle_calls_median,x error plus=error_75, x error minus=error_25, y error plus=oracle_calls_75, y error minus=oracle_calls_25]
  {data/mIQAE-Direct-European_Call_Option-100.dat};
    \addplot[color=gray,domain=5*10e-2:10e-6,style=dashed]{1/(sqrt(x)};
\addlegendentry{mIQAE Square}
\addlegendentry{mIQAE Direct}

\nextgroupplot[title=Put]
    \addplot [color=green,mark=*] 
  plot [error bars/.cd, x dir=both, x explicit, y dir=both, y explicit]
  table [x=error_median,y=oracle_calls_median,x error plus=error_75, x error minus=error_25, y error plus=oracle_calls_75, y error minus=oracle_calls_25]
  {data/mIQAE-Square-European_Put_Option-100.dat};
        \addplot [color=blue,mark=*] 
  plot [error bars/.cd, x dir=both, x explicit, y dir=both, y explicit]
  table [x=error_median,y=oracle_calls_median,x error plus=error_75, x error minus=error_25, y error plus=oracle_calls_75, y error minus=oracle_calls_25]
  {data/mIQAE-Direct-European_Put_Option-100.dat};
    \addplot[color=gray,domain=5*10e-2:10e-6,style=dashed]{1/(sqrt(x)};
    
\nextgroupplot[ylabel=Digital]
    \addplot [color=green,mark=*] 
  plot [error bars/.cd, x dir=both, x explicit, y dir=both, y explicit]
  table [x=error_median,y=oracle_calls_median,x error plus=error_75, x error minus=error_25, y error plus=oracle_calls_75, y error minus=oracle_calls_25] {data/mIQAE-Square-Digital_Call_Option-100.dat};
        \addplot [color=blue,mark=*] 
  plot [error bars/.cd, x dir=both, x explicit, y dir=both, y explicit]
  table [x=error_median,y=oracle_calls_median,x error plus=error_75, x error minus=error_25, y error plus=oracle_calls_75, y error minus=oracle_calls_25]
  {data/mIQAE-Direct-Digital_Call_Option-100.dat};
    \addplot[color=gray,domain=5*10e-2:10e-6,style=dashed]{1/(sqrt(x)};
    
\nextgroupplot
    \addplot [color=green,mark=*] 
  plot [error bars/.cd, x dir=both, x explicit, y dir=both, y explicit]
  table [x=error_median,y=oracle_calls_median,x error plus=error_75, x error minus=error_25, y error plus=oracle_calls_75, y error minus=oracle_calls_25] {data/mIQAE-Square-Digital_Put_Option-100.dat};
        \addplot [color=blue,mark=*] 
  plot [error bars/.cd, x dir=both, x explicit, y dir=both, y explicit]
  table [x=error_median,y=oracle_calls_median,x error plus=error_75, x error minus=error_25, y error plus=oracle_calls_75, y error minus=oracle_calls_25]
  {data/mIQAE-Direct-Digital_Put_Option-100.dat};
    \addplot[color=gray,domain=5*10e-2:10e-6,style=dashed]{1/(sqrt(x)};
\end{groupplot}
\node[anchor=north] (title-x) at ($(group c1r2.south east)!0.5!(group c2r2.south west)-(0,0.5cm)$) {$\epsilon_{\text{QAMC}}$};
\node[anchor=south, rotate=90] (title-y) at ($(group c1r1.south west)!0.5!(group c1r2.north west)-(2,0cm)$) {$N_Q$};

\path (group c1r2.north east) -- node[above=0.5cm]{\ref{groupplot3}} (group c2r2.north west);
\end{tikzpicture}
\caption{Absolute error between the QAMC algorithm and the discretized expectation versus the respective number of calls to the oracle for different precisions $\epsilon$. The dots represent the medians and the error bars the $25$ and $75$ percentiles. Each of the panels corresponds to the payoff of a different option. The experiments have been performed using the square root and direct encodings. }\label{fig:sum_absolute_values_corrected}
\end{figure}

In Figure \ref{fig:sum_absolute_values_corrected} we show the results obtained for the same payoffs as in Figure \ref{fig:probability_loading_true} with both the direct and the square root encoding. As we can see from the figure, the impact of using one or the other encoding is minimal in practice. 

\subsection{Amplitude estimation: mRQAE}\label{sec:new_ae}
In the previous section it was discussed that the discretized expectation can be estimated through the probability of measuring the eigenstate zero of $U_{\text{DE}}$:
\begin{equation*}
    \sqrt{P_{\overline{\ket{0}}}} =\dfrac{1}{\norm{F}}_{\infty}  \left|\sum_{k = 0}^{\mathcal{K}-1}p_S(S_k)F(S_k)\right|.
\end{equation*}
Thus, this partially solves the initial problem. Instead of obtaining the sum of absolute values, something proportional to the absolute value of the sum is returned. Hence, in a situation where the sign of the expectation is of interest, an additional mechanism to overcome this issue is needed. In fact, this is usually the case in financial applications, where the sign makes the difference between a profit and a loss.\\ \\
For this case, we introduce the modified real quantum amplitude estimation (mRQAE) algorithm. The mRQAE is a modified version of the real quantum amplitude estimation (RQAE) (see \cite{rqae}). The main feature of both algorithms is that they are able to read out the size and the sign of the target amplitude. They internally work performing several iterations where, in each iteration they use the Grover amplification algorithm to incrementally amplify the probability of obtaining the target quantum state. The main difference between them is that the number of calls to the amplified state $N_i$, the confidence on each iteration $\gamma_i$ and the required precision on each iteration $\epsilon^p_i$ are chosen in a different manner. In turn, this makes the mRQAE asymptotically more efficient. More precisely, in Table \ref{tab:performance} we show the performance of the mRQAE and the RQAE measured in terms of the number of calls to the oracle $N_Q$ for a given precision $\epsilon$ and confidence $\gamma$ along with the performance of other popular amplitude estimation algorithms in the literature including the aforementioned mIQAE. In Algorithm \ref{alg:mRQAE} there is a schematic description of the code for the mRQAE. For a more thorough revision of the properties of the method we refer the reader to Appendix \ref{appendix:mRQAE}.
\begin{algorithm}[hbtp!]
\tiny
\caption{mRQAE pseudocode.}\label{alg:mRQAE}
\begin{multicols}{2}
\begin{algorithmic}
\State \textbf{Input:}
\Indent
\State $\epsilon$ \Comment{Required precision}
\State $\gamma$ \Comment{$1-\gamma$ is the confidence level}
\State $q$  \Comment{Amplification policy}
\State $\mathcal{A}$ \Comment{Oracle}
\EndIndent
\State \textbf{Output:}
\Indent
\State $a$ \Comment{Estimated amplitude with sign}
\EndIndent
\State \textbf{Algorithm:}
\Indent
\State \Comment{Define relevant parameters}
\State Set $\epsilon^p(q,\infty) = \dfrac{1}{2}\sin^2\left(\dfrac{\pi}{4q}\right)$
\State Set $T = \log_q\left(q^2\dfrac{2\arcsin\left(\sqrt{2\epsilon^p(q,\infty)}\right)}{\arcsin\left(2\epsilon\right)}\right)$
\State Set $k^{\max}=\left\lceil \dfrac{\arcsin\left(\sqrt{2\epsilon^p(q,\infty)}\right)}{\arcsin\left(2\epsilon\right)}-\dfrac{1}{2}\right\rceil$
\State Set $\epsilon^p(q,0) = \dfrac{1}{2}\sin\left(\dfrac{\pi}{2(q+2)}\right)$
\State $i = 1$ \Comment{First Iteration}
\State Set $\gamma_i = \dfrac{\gamma}{2} \dfrac{q-1}{q}\dfrac{1}{2k^{\max}+1}$
\Comment{Confidence for each iteration}\\
\State Set $N_i = \left\lceil \dfrac{1}{2(\epsilon^p(q,0))^2}\log\left(\dfrac{2}{\gamma_i}\right) \right\rceil$ \Comment{Number of shots}
\State Set $\epsilon^p_i = \sqrt{\dfrac{1}{2N_1}\log\left(\dfrac{2}{\gamma_i}\right)}$
\State Set $b = 0.5$\Comment{Shift}
\State Measure $p_{\text{sum}}$ and $p_{\text{diff}}$
\State $a^{\max} = \min\left(\dfrac{\hat{p}_{\text{sum}}-\hat{p}_{\text{diff}}}{4b}+\dfrac{\epsilon^{p}_i}{|2b|},1\right)$
\State $a^{\min} = \max\left(\dfrac{\hat{p}_{\text{sum}}-\hat{p}_{\text{diff}}}{4b}-\dfrac{\epsilon^{p}_i}{|2b|},-1\right)$
\State $a = \dfrac{a^{\max}+a^{\min}}{2}$
\State $\epsilon^a = \dfrac{a^{\max}-a^{\min}}{2}$
\Comment{Following Iterations}
\While{$\epsilon^a>\epsilon$}
\State $i = i+1$
\State Set $b = -a^{\min}$ \Comment{Shift}
\State Set $k = \left\lfloor\dfrac{\pi}{4\arcsin(2\epsilon^a)}-\dfrac{1}{2} \right\rfloor$
\Comment{Number of amplifications}
\State $k = \min(k,k^{\max})$
\State Set $\epsilon^p (q,k) = \dfrac{1}{2}\sin^2\left(\dfrac{\pi}{4\left(q+\dfrac{2}{2k+1}\right)}\right)$
\State Set $\gamma_i = \dfrac{\gamma}{2} \dfrac{q-1}{q}\dfrac{2k + 1}{2k^{\max}+1}$
\Comment{Confidence}\\
\State Set $N_i = \left\lceil \dfrac{1}{2(\epsilon^p(q,k))^2}\log\left(\dfrac{2}{\gamma_i}\right) \right\rceil$ \Comment{Number of shots}
\State Set $\epsilon^p_i = \sqrt{\dfrac{1}{2N_i}\log\left(\dfrac{2}{\gamma_i}\right)}$
\State Measure $p$ \Comment{Shifted probability}
\State \Comment{with $k$ amplifications}
\State $p^{\max}= \min(p+\epsilon^p_i,1)$
\State $p^{\min}= \max(p-\epsilon^p_i,0)$
\State $\theta^{\max} = \dfrac{\arcsin(\sqrt{p^{\max}})}{2k+1}$
\State $\theta^{\min} = \dfrac{\arcsin(\sqrt{p^{\min}})}{2k+1}$
\State $a^{\max} = \sin\left(\theta^{\max}\right)-b $
\State $ a^{\min} =  \sin\left(\theta^{\min}\right)-b$
\State $a = \dfrac{a^{\max}+a^{\min}}{2}$
\State $\epsilon^a = \dfrac{a^{\max}-a^{\min}}{2}$
\EndWhile
\State \textbf{return} $a$
\EndIndent
\end{algorithmic}
\end{multicols}
\end{algorithm}
\begin{table}[htbp!]
    \centering
    \small
    \begin{tabular}{|c|c|}
        \hline
        \cellcolor{g} Algorithm & \cellcolor{g} Performance \\
        \hline
        \textbf{Monte Carlo} & $N_Q^{\text{MC}} = \mathcal{O}\left( \frac{1}{\epsilon^2}\right)$\\
        \hline
        \textbf{QPE}\cite{Brassard_2002} & $N_Q^{\text{QPE}} = \mathcal{O}\left( \frac{1}{\epsilon}\right)$\\
        \hline
        \textbf{MLAE-LIS}\cite{MLAE} & $N_Q^{\text{LIS}} = \mathcal{O}\left(  \frac{1}{\epsilon^{\frac{4}{3}}}\right)$\\
        \hline
        \textbf{MLAE-EIS}\cite{MLAE}& $N_Q^{\text{EIS}} = \mathcal{O}\left(  \frac{1}{\epsilon}\right)$\\
        \hline
        \textbf{PLAE}\cite{giurgicatiron2020low}& $N^{\text{PLAE}}_{\mathcal{A}} = \mathcal{O}\left( \frac{1}{\epsilon^{1+\beta}}\right)$, $d = \mathcal{O}\left( \frac{1}{\epsilon^{1-\beta}}\right)$\\
        \hline
        \textbf{Improved MLAE}\cite{jittering}& $N_Q^{\text{impMLAE}} = \mathcal{O}\left( \frac{1}{\epsilon} \frac{1}{d} \log(\frac{1}{\gamma})\right)$, $d=2^{q-2}$\\
        \hline
        \textbf{IQAE} \cite{Grinko_2021}& $N_Q^{\text{IQAE}} <  \frac{50}{\epsilon} \log{\left( \frac{2}{\gamma} \log_2{\frac{\pi}{4\epsilon}} \right)}$\\
        \hline
        \textbf{mIQAE}\cite{Fukuzawa_2023} & $N_Q^{\text{mIQAE}} < \frac{123}{\epsilon} \log{\frac{6}{\gamma}}$\\
        \hline
        \textbf{QCoin} \cite{Abrams1999FastQA}& $N_Q^{\text{QCoin}} = \mathcal{O}\left(  \frac{1}{a} \frac{1}{\epsilon} \log{\frac{1}{\gamma}}\right)$ , $k \geq 2$, $1 \geq q \geq (k-1)$\\
        \hline
        \textbf{QoPrime} \cite{giurgicatiron2020low}& $N_Q^{\text{QoPrime}} < C \lceil\frac{k}{q} \rceil \frac{1}{\epsilon^{1+q/k}} \log\left(\frac{4}{\gamma} \lceil\frac{k}{q} \rceil \right)$, $d = \mathcal{O}\left( \frac{1}{\epsilon^{1-q/k}}\right)$ \\
        \hline
        \textbf{FasterAE} \cite{faster_AE}& $N_Q^{\text{fasterAE}} <  \frac{4.1 \cdot 10^3}{\epsilon}\log{\left( \frac{4}{\gamma} \log_2 \left(\frac{2 \pi }{3\epsilon}\right)\right)}$ \\
        \hline
        \textbf{AdaptiveAE} \cite{adaptive_QAE}& $N_Q^{\text{adaptiveAE}} < \mathcal{O}\left(\frac{1}{\epsilon} \log{\left( \frac{\pi^2(T+1)}{3\gamma} \right)}\right)$, $T = \lceil \frac{\log \frac{\pi}{K\epsilon}}{\log K} \rceil$\\
        \hline
        \textbf{RQAE} \cite{rqae}& $N_Q^{\text{RQAE}} < \frac{C_1(q)}{\epsilon_a} \log\left[ \frac{3.3}{\gamma} \log_q \left( \frac{C_2(q)}{\epsilon}\right)\right]$\\
        \hline
        \textbf{mRQAE} \eqref{eq:mrqae_oracle_bound}& $N_Q^{\text{MRQAE}} < \frac{C_1(q)}{\epsilon} \log\left[ \frac{C_2(q)}{\gamma} \right]$\\
        \hline
    \end{tabular}
    \caption{Performance of different amplitude estimation algorithms. $N_Q$ denotes the number of calls to the oracle, $\epsilon$ is the target precision and $1-\gamma$ is the confidence level. Other parameters appearing in the table are related to each specific algorithm. For a full description of their meaning the reader is referred to the associated references. The $\sim$ symbol indicates that the algorithm has an asymptotic behaviour, while the $<$ indicates that the performance is proved rigorously. }
    \label{tab:performance}
\end{table}

\begin{figure}[htbp!]
\centering
\begin{tikzpicture}
\begin{axis}[
xmode = log,
ymode = log,
xmin=10e-7,
xmax=0.01,
max space between ticks=50pt,
x dir=reverse,
xlabel=$\epsilon_{\text{QAMC}}$,
ylabel=$N_Q$,
legend pos=north west,
]\addplot [color=green,mark=*] 
  plot [error bars/.cd, x dir=both, x explicit, y dir=both, y explicit]
  table [x=error_median,y=oracle_calls_median,x error plus=error_75, x error minus=error_25, y error plus=oracle_calls_75, y error minus=oracle_calls_25] {data/mIQAE-Square-Parts-Futures-100.dat};
  \addplot [color=blue,mark=*] 
  plot [error bars/.cd, x dir=both, x explicit, y dir=both, y explicit]
  table [x=error_median,y=oracle_calls_median,x error plus=error_75, x error minus=error_25, y error plus=oracle_calls_75, y error minus=oracle_calls_25] {data/mIQAE-Direct-Parts-Futures-100.dat};
  \addplot [color=red,mark=*] 
  plot [error bars/.cd, x dir=both, x explicit, y dir=both, y explicit]
  table [x=error_median,y=oracle_calls_median,x error plus=error_75, x error minus=error_25, y error plus=oracle_calls_75, y error minus=oracle_calls_25] {data/mRQAE-Direct-Futures-None.dat};
  \addlegendentry{mIQAE Square}
  \addlegendentry{mIQAE Direct}
  \addlegendentry{mRQAE Direct}
\end{axis}
\end{tikzpicture}
\caption{Absolute error between the QAMC algorithm and the discretized expectation of an option with a payoff $(S_T-K)$ with $K = 1.5 S_t$ versus number of calls to the oracle for different values of precision $\epsilon$. The dots represent the medians and the error bars the $25$ and $75$ percentiles. For the mRQAE we have used the direct encoding and for the mIQAE we have applied both techniques. Moreover, in the case of the mIQAE we have separated the negative and positive parts of the payoff for a correct pricing.}\label{fig:sum_absolute_values_RQAE}
\end{figure}
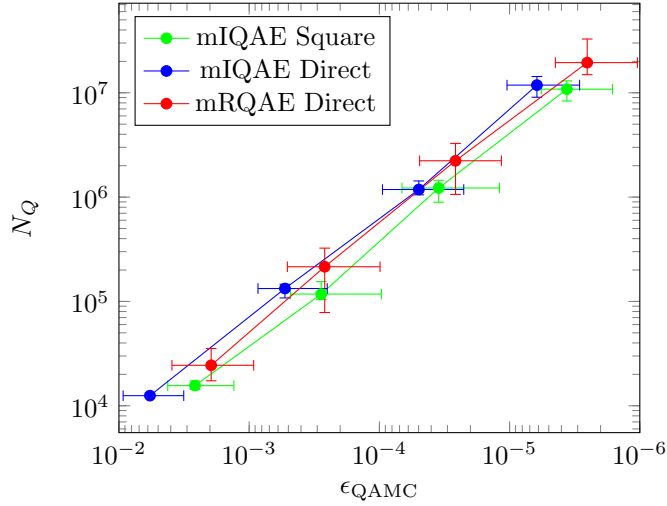

 In Figure \ref{fig:sum_absolute_values_RQAE}, an example where the price of the derivative becomes negative is shown. As it is shown, mRQAE is able to recover the true price without requiring additional mechanisms. Moreover, it is competitive with the mIQAE which is currently considered one of the most efficient algorithms in the literature in terms of number of calls to the oracle.

\begin{figure}[htbp!]
    \centering
\begin{tikzpicture}
\begin{groupplot}[group style={
group size=2 by 2,
vertical sep=2cm,horizontal sep=1cm},
width=0.4\textwidth,
xmode = log,
ymode = log,
xmin=10e-8,
xmax=0.05,
ymin=10e2,
ymax=5*10e6,
x dir=reverse,
]
\nextgroupplot[title=Call,legend to name={groupplot4},ylabel=European,legend style={legend columns=3}]
    \addplot [color=green,mark=*] 
  plot [error bars/.cd, x dir=both, x explicit, y dir=both, y explicit]
  table [x=error_median,y=oracle_calls_median,x error plus=error_75, x error minus=error_25, y error plus=oracle_calls_75, y error minus=oracle_calls_25]
  {data/mIQAE-Square-European_Call_Option-100.dat};
      \addplot [color=blue,mark=*] 
  plot [error bars/.cd, x dir=both, x explicit, y dir=both, y explicit]
  table [x=error_median,y=oracle_calls_median,x error plus=error_75, x error minus=error_25, y error plus=oracle_calls_75, y error minus=oracle_calls_25]
  {data/mIQAE-Direct-European_Call_Option-100.dat};
  plot [error bars/.cd, x dir=both, x explicit, y dir=both, y explicit]
  \addplot [color=red,mark=*]
  table [x=error_median,y=oracle_calls_median,x error plus=error_75, x error minus=error_25, y error plus=oracle_calls_75, y error minus=oracle_calls_25]
  {data/mRQAE-Direct-European_Call_Option-None.dat};
\addlegendentry{mIQAE Square}
\addlegendentry{mIQAE Direct}
\addlegendentry{mRQAE Direct}

\nextgroupplot[title=Put]
    \addplot [color=green,mark=*] 
  plot [error bars/.cd, x dir=both, x explicit, y dir=both, y explicit]
  table [x=error_median,y=oracle_calls_median,x error plus=error_75, x error minus=error_25, y error plus=oracle_calls_75, y error minus=oracle_calls_25]
  {data/mIQAE-Square-European_Put_Option-100.dat};
        \addplot [color=blue,mark=*] 
  plot [error bars/.cd, x dir=both, x explicit, y dir=both, y explicit]
  table [x=error_median,y=oracle_calls_median,x error plus=error_75, x error minus=error_25, y error plus=oracle_calls_75, y error minus=oracle_calls_25]
  {data/mIQAE-Direct-European_Put_Option-100.dat};
  \addplot [color=red,mark=*]
  table [x=error_median,y=oracle_calls_median,x error plus=error_75, x error minus=error_25, y error plus=oracle_calls_75, y error minus=oracle_calls_25]
  {data/mRQAE-Direct-European_Put_Option-None.dat};
    
\nextgroupplot[ylabel=Digital]
    \addplot [color=green,mark=*] 
  plot [error bars/.cd, x dir=both, x explicit, y dir=both, y explicit]
  table [x=error_median,y=oracle_calls_median,x error plus=error_75, x error minus=error_25, y error plus=oracle_calls_75, y error minus=oracle_calls_25] {data/mIQAE-Square-Digital_Call_Option-100.dat};
        \addplot [color=blue,mark=*] 
  plot [error bars/.cd, x dir=both, x explicit, y dir=both, y explicit]
  table [x=error_median,y=oracle_calls_median,x error plus=error_75, x error minus=error_25, y error plus=oracle_calls_75, y error minus=oracle_calls_25]
  {data/mIQAE-Direct-Digital_Call_Option-100.dat};
  \addplot [color=red,mark=*]
  table [x=error_median,y=oracle_calls_median,x error plus=error_75, x error minus=error_25, y error plus=oracle_calls_75, y error minus=oracle_calls_25]
  {data/mRQAE-Direct-Digital_Call_Option-None.dat};
    
\nextgroupplot
    \addplot [color=green,mark=*] 
  plot [error bars/.cd, x dir=both, x explicit, y dir=both, y explicit]
  table [x=error_median,y=oracle_calls_median,x error plus=error_75, x error minus=error_25, y error plus=oracle_calls_75, y error minus=oracle_calls_25] {data/mIQAE-Square-Digital_Put_Option-100.dat};
        \addplot [color=blue,mark=*] 
  plot [error bars/.cd, x dir=both, x explicit, y dir=both, y explicit]
  table [x=error_median,y=oracle_calls_median,x error plus=error_75, x error minus=error_25, y error plus=oracle_calls_75, y error minus=oracle_calls_25]
  {data/mIQAE-Direct-Digital_Put_Option-100.dat};
  \addplot [color=red,mark=*]
  table [x=error_median,y=oracle_calls_median,x error plus=error_75, x error minus=error_25, y error plus=oracle_calls_75, y error minus=oracle_calls_25]
  {data/mRQAE-Direct-Digital_Put_Option-None.dat};
\end{groupplot}
\node[anchor=north] (title-x) at ($(group c1r2.south east)!0.5!(group c2r2.south west)-(0,0.5cm)$) {$\epsilon_{\text{QAMC}}$};
\node[anchor=south, rotate=90] (title-y) at ($(group c1r1.south west)!0.5!(group c1r2.north west)-(2,0cm)$) {$N_Q$};

\path (group c1r2.north east) -- node[above=0.5cm]{\ref{groupplot4}} (group c2r2.north west);
\end{tikzpicture}
\caption{Absolute error between the QAMC algorithm and the discretized expectation versus the respective number of calls to the oracle for different precisions $\epsilon$. The dots represent the medians and the error bars the $25$ and $75$ percentiles. Each of the panels corresponds the payoff of a different option. For the mRQAE we have used the direct encoding and for the mIQAE we have applied both techniques.}\label{fig:RQAE_speed_comparison}
\end{figure}
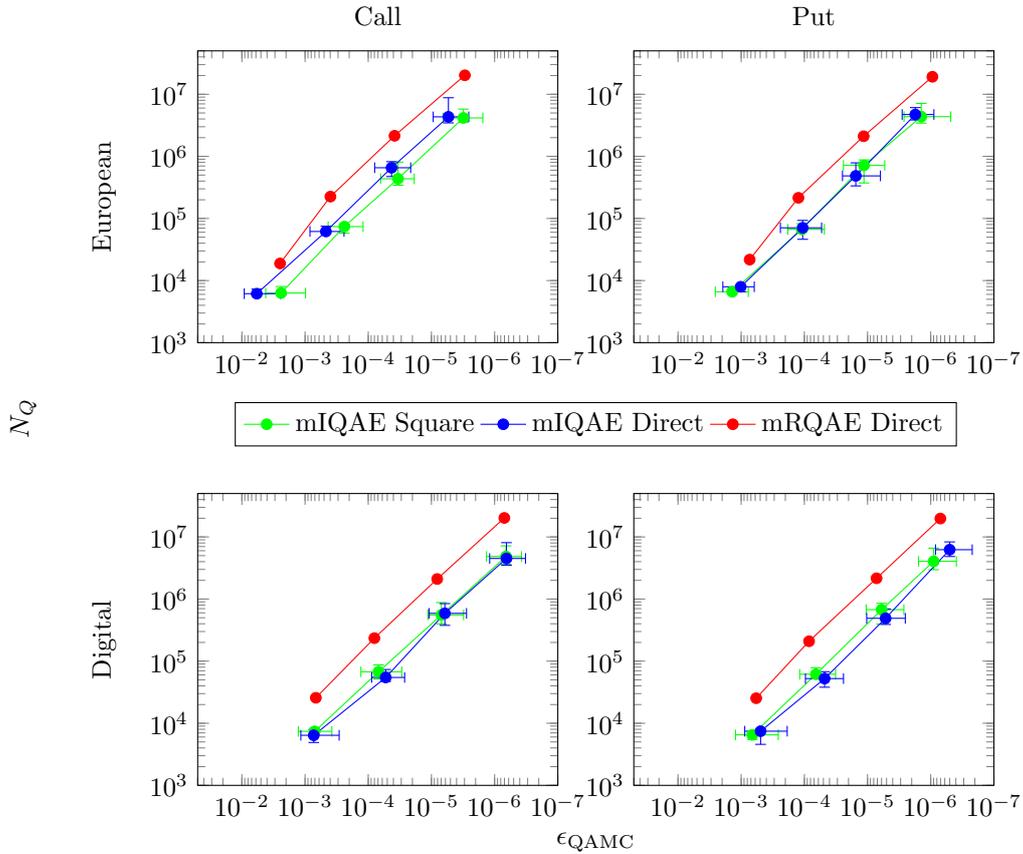

 In Figure \ref{fig:RQAE_speed_comparison} we show that the mRQAE obtains a reasonable performance compared with the mIQAE when we use it for contracts with positive payoff.

\subsection{Computer implementation details}
	Concerning the hardware, all the tests have been performed  the Finisterrae III (FT3) infraestructure provided by CESGA. On the software side, we have mainly used our \textit{Python} library \cite{NEASQC_library} which is built on top of the QLM library provided by Atos. In Table \ref{tab:configuration} we provide further details.

 \begin{table}[htbp!]
    \centering
    \small
    \begin{tabular}{|c|c|c|}
        \hline
        \cellcolor{g} Provider & CESGA & CESGA\\
        \hline
        \cellcolor{g} Node & QLM QAPTIVA & ILK\\
        \hline
        \cellcolor{g} CPU & 48 Intel(R) Xeon(R) Platinum 8260L & 2 Intel Xeon Ice Lake 8352Y\\
        \hline
        \cellcolor{g} RAM & 1510 GB & 256GB\\
        \hline
        \cellcolor{g} OS & Red Hat Enterprise Linux release 9.2 & Rocky Linux release 8.4\\
        \hline
        \cellcolor{g} Python version & 3.9.16 & 3.9.9\\
        \hline 
        \cellcolor{g} QLM version & QLM-1.9.1 & myQLM-1.9.9\\
        \hline
    \end{tabular}
    \caption{Hardware and software configuration for the experiments.}
    \label{tab:configuration}
\end{table}

\section{Conclusions}\label{sec:conclusions}
In this article, we have presented a novel proposals to leverage quantum computers for derivative pricing. This proposal combines the direct encoding protocol with the mRQAE algorithm to price derivative products with
negative payoffs without the need of separating the payoff function into positive and negative codomains, as it is mandatory in the standard QAMC algorithm to obtain correct prices. Moreover, we have performed several experiments to validate our methodology and showed that the new methodology is competitive in terms of “speed” with the current standards.\\

The direct encoding introduces an alternative way to perform QAMC. It is somehow more natural than its predecessors, since we directly work with the payoffs and not with their square roots. However, in order to obtain the most of it we need to combine it with algorithms such as the mRQAE. The mRQAE is capable of reading out the sign of the underlying amplitude and offers some control over the depth of the circuits, a crucial feature in the current NISQ era.\\

Although in theory QAMC exhibits a quadratic speedup over CMC, there are still many issues to solve
in practice, specially if we take into consideration the current state-of-the-art hardware constraints.
First, as explained in Section \ref{sec:standard_encoding}, the implementation of the oracle $U_S$ requires an excessively large number of qubits. Second, the depths required by the current Grover-like routines are not feasible under the current decoherence times. Finally, the total number of gates when combining the implementation
of oracle $U_S$ with a Grover-like algorithm requires a gate error beyond the capabilities of the current
technology.\\

The possible extensions or improvements of the proposed techniques are all centered around addressing the previously pointed problems. Here we mention three areas that might be of broad interest for the QAMC community. \\

First, although we have seen that it is theoretically possible to efficiently initialize a state through the direct simulation of a target SDE, it is too demanding in terms of hardware requirements. For this reason, we need to consider new ways to efficiently initialize states that approximately encode the target probability distribution. This is the main step towards a real end to end implementation of the QAMC.\\

A second research area aims to investigate the proposal of new encodings. Both the direct and the square encodings suffer from the fact that the payoff needs to be normalized. Otherwise, we need to define an upper bound for the payoff, which induces a truncation error. Other encodings could work with different normalizations, potentially reducing the effects of truncation.\\

Third, we point towards possible extensions of the mRQAE. The choice of parameters considered in the present paper is not unique and alternative choices could be more efficient in terms, for example, of the total number of shots. In fact, we believe that it is interesting to explore different parameter settings in the future.

\section{Declarations}
    \subsection{Availability of data and materials}
    
        Not applicable.

    \subsection{Competing interests}
    
        The authors declare that they have no competing interests.
        
    \subsection{Funding}
    
        All  authors  acknowledge  the  European  Project  NExt  ApplicationS  of  Quantum  Computing (NEASQC), funded by Horizon 2020 Program inside the call H2020-FETFLAG-2020-01(Grant Agreement 951821). Á. Leitao, A. Manzano and C. Vázquez also acknowledge funding from the Galician Government (grant ED431C 2022/47, including FEDER financial support). Á. Leitao, A. Manzano and C. Vázquez also acknowledge the support from CITIC, as a center accredited for excellence within the Galician University System and a member of the CIGUS Network, receives subsidies from the Department of Education, Science, Universities, and Vocational Training of the Xunta de Galicia. Additionally, it is co-financed by the EU through the FEDER Galicia 2021-27 operational program (Ref. ED431G 2023/01).

    \subsection{Authors' contributions}
    A.M. wrote the main manuscript.
    G.F. performed the experiments.
    All authors provided ideas related to the different aspects of the work (financial and quantum computing theoretical, practical and implementation issues) and reviewed the manuscript.

    \subsection{Acknowledgements}

The authors would like to thank  Vedran Dunjko and Emil Dimitrov for fruitful discussions on some aspects of the present work.
    
The authors also acknowledge Galicia Supercomputing Center (CESGA) for providing access to FinisTerrae III supercomputer with financing from the Programa Operativo Plurirregional de España 2014-2020 of ERDF, ICTS-2019-02-CESGA-3.\\

%% file: body/appendix1.tex
\section{Some useful inequalities}
\label{ch:appendix1}
In Appendix A we prove several lemmas 
that state some inequalities that are used in this article.

\begin{lemma}\label{lemma:upper_bound_sine}
    For all $a,b\in [0,1]$ such that $a>b$, we have that:
    \begin{equation*}
        \sin(a)-\sin(b)\leq \sin(a-b).
    \end{equation*}
\end{lemma}
\begin{proof}
In order to prove it, we will show that the function:
\begin{equation}
    f(a,b) = \sin(a)-\sin(b)- \sin(a-b),
\end{equation}
is less or equal than zero for the region delimited by
\begin{equation}
\begin{aligned}
    & b = 0,\, a\in[0,1]\, \text{marked in blue},\\
    & a=b,\, a,b\in[0,1]\, \text{marked in green},\\
    & a = 1,\, b\in[0,1]\, \text{marked in red}.\\
\end{aligned}
\qquad\qquad
\vcenter{\hbox{
\begin{minipage}{5cm}
\centering
\begin{tikzpicture}[scale=1.]
    \coordinate (zero) at ( $ ({0},{0}) $ );
    \coordinate (a_max) at ( $ ({2.},{0}) $ );
    \coordinate (b_max) at ( $ ({0},{2.}) $ );
    \coordinate (shift_up) at ( $ ({0},{0.3}) $ );
    \coordinate (shift_down) at ( $ ({0},{-0.3}) $ );
    \coordinate (shift_right) at ( $ ({0.3},{0}) $ );
    \coordinate (shift_left) at ( $ ({-0.3},{0}) $ );
    \draw[->,ultra thick,opacity=0.3] (0,0)--(3,0) node[right]{$a$};
    \draw[->,ultra thick,opacity=0.3] (0,0)--(0,3) node[above]{$b$};
    \node[] at ($ (zero)+(-0.3,-0.3)$) {0};
    \draw[dashed,opacity=0.3] (b_max)--($(a_max)+(b_max)$);
    \draw [pattern=north west lines, pattern color=black]
       (zero) -- (a_max) -- ($ (a_max)+(b_max) $) -- cycle;
    \draw[-,color=blue,ultra thick] ($ (zero)$) -- ($ (a_max) $);
    \node[] at ($ (a_max)+(shift_down)$) {1};
    \draw[-,color=red,ultra thick] ($ (a_max)$) -- ($ (b_max)+(a_max)$);
    \node[] at ($ (b_max)+(shift_left)$) {1};
    \draw[-,color=green,ultra thick] ($ (zero)$) -- ($ (b_max)+(a_max)$);
\end{tikzpicture}
\end{minipage}}}
\end{equation}
First we show that $f\leq 0$ on the boundaries:
\begin{itemize}
    \item In the blue region $b = 0$ we have that $f = 0$.
    \item In the green region $a = b$ we have that $f = 0$.
    \item In the red region we have that $f(1,b) = \sin(1)-\sin(b)-\sin(1-b)$. In this region it is not obvious that the function is smaller than zero. However, it is easy to verify that $f(1,b)$ has a global minimum at $b = 0.5$ which is smaller than zero. In the region $b\in[0,1/2)$ is monotonically decreasing and in the region $b\in(1/2,1)$ is monotonically increasing with $f(1,1) = 0$, so that $f(1,b)\leq 0$.
\end{itemize}
In the interior it is enough to notice that $\dfrac{\partial f}{\partial a}\leq 0\, \forall a\in (0,1)$. If we start in the green boundary where the function is equal to zero and we move along the $a$ axis, the function strictly decreases. Therefore, it must be smaller of equal than zero.\\ \\
As we have proven that $f\leq 0$ in the interior plus the boundaries we have proven the inequality.
\end{proof}

\begin{lemma}\label{lemma:lower_bound_sine}
    For all $a,b\in [0,1]$ $a>b$, we have that:
    \begin{equation*}
        \sin(a)-\sin(b)\geq \sin\left(\dfrac{a-b}{2}\right).
    \end{equation*}
\end{lemma}
\begin{proof}
In order to prove it, we will show that the function:
\begin{equation}\label{eq:sina-sinb-sina(a-b)}
    f(a,b) = \sin(a)-\sin(b)-\sin\left(\dfrac{a-b}{2}\right),
\end{equation}
is greater than zero for the region delimited by
\begin{equation}
\begin{aligned}
    & b = 0,\, a\in[0,1]\, \text{marked in blue},\\
    & a=b,\, a,b\in[0,1]\, \text{marked in green},\\
    & a = 1,\, b\in[0,1]\, \text{marked in red}.\\
\end{aligned}
\qquad\qquad
\vcenter{\hbox{
\begin{minipage}{5cm}
\centering
\begin{tikzpicture}[scale=1.]
    \coordinate (zero) at ( $ ({0},{0}) $ );
    \coordinate (a_max) at ( $ ({2.},{0}) $ );
    \coordinate (b_max) at ( $ ({0},{2.}) $ );
    \coordinate (shift_up) at ( $ ({0},{0.3}) $ );
    \coordinate (shift_down) at ( $ ({0},{-0.3}) $ );
    \coordinate (shift_right) at ( $ ({0.3},{0}) $ );
    \coordinate (shift_left) at ( $ ({-0.3},{0}) $ );
    \draw[->,ultra thick,opacity=0.3] (0,0)--(3,0) node[right]{$a$};
    \draw[->,ultra thick,opacity=0.3] (0,0)--(0,3) node[above]{$b$};
    \node[] at ($ (zero)+(-0.3,-0.3)$) {0};
    \draw[dashed,opacity=0.3] (b_max)--($(a_max)+(b_max)$);
    \draw [pattern=north west lines, pattern color=black]
       (zero) -- (a_max) -- ($ (a_max)+(b_max) $) -- cycle;
    \draw[-,color=blue,ultra thick] ($ (zero)$) -- ($ (a_max) $);
    \node[] at ($ (a_max)+(shift_down)$) {1};
    \draw[-,color=red,ultra thick] ($ (a_max)$) -- ($ (b_max)+(a_max)$);
    \node[] at ($ (b_max)+(shift_left)$) {1};
    \draw[-,color=green,ultra thick] ($ (zero)$) -- ($ (b_max)+(a_max)$);
\end{tikzpicture}
\end{minipage}}}
\end{equation}
First we use a trigonometric identity on the sum of sine functions and transform Equation \eqref{eq:sina-sinb-sina(a-b)} into:
\begin{equation}\label{eq:f(a,b)}
\begin{aligned}
    f(a,b) &= 2\cos\left(\dfrac{a+b}{2}\right)\sin\left(\dfrac{a-b}{2}\right)-\sin\left(\dfrac{a-b}{2}\right)\\
    &= \left(2\cos\left(\dfrac{a+b}{2}\right)-1\right)\sin\left(\dfrac{a-b}{2}\right).
    \end{aligned}
\end{equation}
Next, we show that the two factors in the product are positive. For the first term:
\begin{equation}
    \sin\left(\dfrac{a-b}{2}\right),
\end{equation}
this is straightforward. The second term:
\begin{equation*}
    \left(2\cos\left(\dfrac{a+b}{2}\right)-1\right)
\end{equation*}
has a global minimum $2\cos(1) - 1$ at $a = b = 1$.

Therefore, since both factors are positive in the prescribed region, the product is positive.
\end{proof}

\begin{lemma}\label{lemma:upper_bound_arcsin}
    For all $a,b\in [0,\arcsin(1)]$ such that $a>b$, we have that:
    \begin{equation*}
        \arcsin(a)-\arcsin(b)\leq 2 \arcsin(a-b).
    \end{equation*}
\end{lemma}
\begin{proof}
From Lemma \ref{lemma:lower_bound_sine} we know that for $\alpha,\beta\in [0,1]$ we have that:
\begin{equation*}
    \sin(\alpha)-\sin(\beta)\geq \sin\left(\dfrac{\alpha-\beta}{2}\right).
\end{equation*}
Now we define the two variables $a = \arcsin(\alpha)$ and $b = \arcsin(\beta)$ and substitute them into the previous equation getting:
\begin{equation*}
    a-b\geq \sin\left(\dfrac{\arcsin{a}-\arcsin{b}}{2}\right).
\end{equation*}
Taking the arcsin on both sides we get the desired identity. Note that this can be done since the arcsin in the defined interval is a positive increasing function.
\end{proof}

\begin{lemma}\label{lemma:upper_bound_square_root}
    For all $a,b\in [0,1]$ such that $a>b$, we have that:
    \begin{equation*}
        \sqrt{a}-\sqrt{b}\leq \sqrt{a-b}.
    \end{equation*}
\end{lemma}
\begin{proof}
If we assume that the stated inequality does not hold, by 
taking the square on both sides of the resulting expression, we get:
\begin{equation*}
    a+b-2\sqrt{ab} > a-b \implies b > \sqrt{a b} \implies \sqrt{b} > \sqrt{a} \implies b > a,
\end{equation*}
which contradicts the hypothesis $a>b$.
\end{proof}

The following lemma is a modification of Lemma 3.4 in \cite{Fukuzawa_2023}:
\begin{lemma}\label{lemma:upper_bound_series}
    Consider an increasing sequence $x_0 = 1,x_1,...x_N$ such that $x_N\leq x_{\max}$ and $x_{i+1}\geq q x_i\; \forall i \in \{0,...,N-1\},\, q>1$. Moreover, suppose that $f$ is a positive increasing function over the range $[1,x_{\max}]$, then:
    \begin{equation}
        \sum_{i = 1}^Nf(x_i)\leq \sum_{i = 0}^{N-1}f\left(\dfrac{x_{\max}}{q^i}\right).
    \end{equation}
\end{lemma}
\begin{proof}
First note that:
\begin{equation*}
    x_j\leq \dfrac{x_N}{q^{N-j}}\leq \dfrac{x_{\max}}{q^{N-j}}.
\end{equation*}
Since $f$ is a positive increasing function,
\begin{equation*}
    f(x_j)\leq f\left(\dfrac{x_{\max}}{q^{N-j}}\right)\implies \sum_{j = 1}^N f(x_j) \leq \sum_{j = 1}^N f\left(\dfrac{x_{\max}}{q^{N-j}}\right) = \sum_{j = 0}^{N-1} f\left(\dfrac{x_{\max}}{q^{N-j}}\right).
\end{equation*}
\end{proof}

%% file: body/appendix2.tex
\section{Modified Real Quantum Amplitude Estimation}\label{appendix:mRQAE}

\begin{figure}[hbtp!]
        \centering
    \begin{subfigure}[t]{0.3\textwidth}
        \centering
        \includegraphics[width=\textwidth]{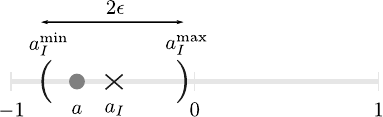}
    \end{subfigure}
     \hfill
    \begin{subfigure}[t]{0.3\textwidth}
        \centering
        \includegraphics[width=\textwidth]{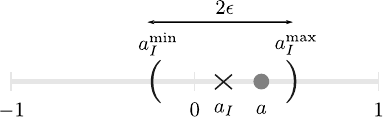}
    \end{subfigure}
     \hfill
    \begin{subfigure}[t]{0.3\textwidth}
        \centering
        \includegraphics[width=\textwidth]{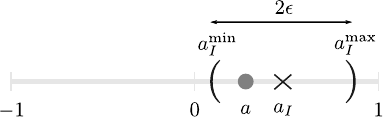}
    \end{subfigure}
    \caption{Illustration of the three positions of the confidence interval with respect to zero. The different symbols include $a$ which is the target amplitude to be estimated, $2\epsilon$ which is the width of the confidence interval with bounds $(a_I^{\min},a_I^{\max})$ and $a_I$ which is the center of the confidence interval.}
    \label{fig:initial_problem}
\end{figure}

Consider a one-parameter family of oracles $\mathcal{A}_{b}$ that, acting on the state $\ket{0}$, yield
\begin{equation}\label{eq:shi_def}
    \mathcal{A}_{b}\ket{0} = \ket{\psi} = \left(a+b\right)\ket{\phi}+c_b\, \ket{\phi^\perp}_b\ ,
\end{equation}
where $a$ is a real number, $b$ is an auxiliary, continuous and real parameter that we call ``shift'', and $\ket{\phi}$ is a specified direction in the Hilbert space ${\cal H}$.
The mRQAE algorithm estimates the amplitude $a$ exploiting the possibility of tuning the shift $b$ iteratively. The ket $\ket{\psi}$ belongs to the plane $\Pi_b = \text{span}\{\ket{\phi},\ket{\phi^\perp}_b\} \subset {\cal H}$ for which the kets $\ket{\phi}$ and $\ket{\phi^\perp}_b$ provide an orthonormal basis. Note that all the quantities with a sub-index $b$ depend on the actual value of the shift. In practice, the construction of oracles such as $\mathcal{A}_b$ from a given un-shifted oracle $\mathcal{A}$ is generally not difficult. In most cases, a controlled shift of an amplitude can be efficiently implemented via Hadamard gates and some controlled operations. In particular, its implementation is straightforward in the framework described in \cite{QNP}.\\ \\ 
Given a precision level $\epsilon$ and a confidence level $1-\gamma$, the goal of the algorithm is to compute a confidence interval $(a_I^{\min},a_I^{\max}) \subset [-1,1]$ of width smaller than $2\epsilon$ which contains the value of $a$ with probability greater or equal to $1-\gamma$ (see Figure \ref{fig:initial_problem}). Here, $I$ denotes the number of iterations to achieve the prescribed accuracy. We take as a representative of the interval its center, $a_I = \frac{a_I^{\min}+a_I^{\max}}{2}$, thus admitting a maximum error of $\epsilon$:
\begin{equation}\label{eq:initial_problem}
        \mathbb{P}\Big[|a-a_I|\geq \epsilon \Big] \leq \gamma\ .
\end{equation}

It is convenient to express the amplitudes in terms of their corresponding angles by means of the generic mapping $\theta_x = \arcsin(x)$ for any real amplitude $x$. Note that the angle representation is particularly suited to describe Grover amplifications, which indeed admit an interpretation as rotations in the plane $\Pi_b$. As an example, the state $\ket{\psi}$ given in \eqref{eq:shi_def} can be written as 
\begin{equation}
    \ket{\psi} = \sin(\theta_{a+b})\ket{\phi}+\cos(\theta_{a+b})\ket{\phi^\perp}_b\ ,
\end{equation}
where $\theta_{a+b}$ represents a rotation in the plane $\Pi_b$ defined above.
Throughout this chapter, we will be changing back and forth from the representation in terms of the actual amplitude or its associated angle whenever needed. In order to avoid notational clutter, we henceforth drop the sub-index $b$ on the perpendicular ket, leaving its dependence on the shift as understood. Actually, such dependence does not play any role for the algorithm.
\\

In the following subsections we describe step by step the inner workings of the mRQAE.

\subsection{First iteration: estimating the sign}\label{sign}
This step achieves a first estimation of the bounds of the confidence interval $(a_1^{\min},a_1^{\max})$.  Normally, this estimation would not be sensitive to the sign of the underlying amplitude because, when sampling from a quantum state, we obtain the square of the amplitude. Nevertheless, by taking advantage of the shift $b$ we can circumvent this limitation. In order to compute the sign, we will combine two different pieces of information: the result of measuring the two oppositely shifted states $\ket{\psi_1}_{\pm}$ defined as:
\begin{equation}\label{mas}
\begin{aligned}
    &\ket{\psi_1}_{-}
    := \left(a-b_1\right) \ket{\phi} + ...\
    &\ket{\psi_1}_{+}
    :=
    \left(a+b_1\right)\ket{\phi} + ... \ , 
     ,
\end{aligned}
\end{equation}
for an arbitrary real shift $b_1$. The sign of $b_1$ has to be decided at the start of the algorithm to have a clear reference. In practice, in some setups it is possible to measure at the same time both states taking advantages of Hadamard gates as in the quantum arithmetic techniques discussed in \cite{QNP}.
As $a$ and $b_1$ are real numbers, we have the identity: 
\begin{equation}\label{ide}
     a = \dfrac{\left(a+b_1\right)^2-\left(a-b_1\right)^2}{4b_1}\ ,
\end{equation}
and we can build a first empirical estimation $\hat{a}_1$ of $a$ as follows:
\begin{equation}\label{eqn:first_empirical_amplitude_estimation}
    \hat{a}_1 = \dfrac{\hat{p}_{\text{sum}}-\hat{p}_{\text{diff}}}{4b_1}\ ,
\end{equation}
where $\hat{p}_{\text{sum}}$ and $\hat{p}_{\text{diff}}$ are the empirical probabilities of getting $\ket{\phi}$ when measuring $\ket{\psi_1}_{-}$ and $\ket{\psi_1}_{+}$, respectively. Throughout this chapter, when we measure, we will use $\hat{p}$ to denote the empirical probability obtained from direct sampling. As an example, if in iteration $i$ we sample the state $N_i$ times, getting $\ket{\phi}$ as an outcome $\hat{N}_i$ times, the estimated probability of $\ket{\phi}$ will be $\hat{p}_i = \dfrac{\hat{N}_i}{N_i}$.\\

\begin{figure}[hbtp!]
        \centering
    \begin{subfigure}[c]{0.45\textwidth}
        \centering
        \includegraphics[width=\textwidth]{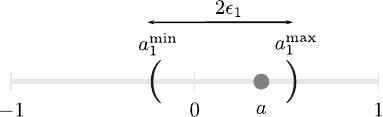}
        \label{fig:amplitude_problem_image}
    \end{subfigure}
        \hfill
    \begin{subfigure}[c]{0.45\textwidth}
        \centering
        \includegraphics[width=\textwidth]{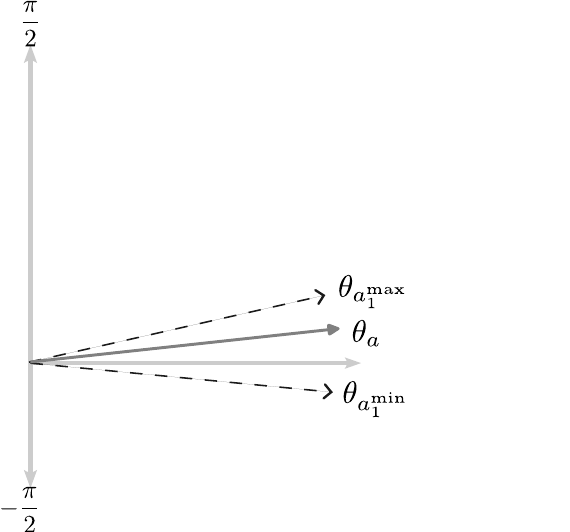}
        \label{fig:angle_problem_image}
    \end{subfigure}
    \caption{First confidence interval in terms of amplitudes an angles. In the figure on the left, the dot corresponds to $a$, namely the probability to be estimated; $a_1^{\min}$ and $a_1^{\max}$ define the confidence interval whose width is $2\epsilon_1$. In the figure on the right, the same confidence interval is represented in terms of angles. Note that the ``true value" (represented by either $a$ and $\theta_a$) falls inside the confidence interval. In order to avoid clutter, here we are not representing the central value of the confidence interval.}
    \label{fig:angle_problem}
\end{figure}

From \eqref{ide} and \eqref{eqn:first_empirical_amplitude_estimation}, we can obtain a first confidence interval $(a_1^{\min},a_1^{\max})$, with:
\begin{equation}\label{eq:first_amplitude_bounds}
      \begin{aligned}
    &a^{\min}_1 = \max\left(\dfrac{\hat{p}_{\text{sum}}-\hat{p}_{\text{diff}}}{4b_1}-\dfrac{\epsilon^{p}_1}{|2b_1|},-1\right),
    &a^{\max}_1 = \min\left(\dfrac{\hat{p}_{\text{sum}}-\hat{p}_{\text{diff}}}{4b_1}+\dfrac{\epsilon^{p}_1}{|2b_1|},1\right),
    \end{aligned} 
\end{equation}
so that,
\begin{equation}
\begin{aligned}
    &a_1 = \dfrac{a^{\max}_1+a^{\min}_1}{2}\ , 
    &\epsilon_1^a = \dfrac{a^{\max}_1-a^{\min}_1}{2}\ ,
\end{aligned} 
\end{equation}
where the $\max$ and the $\min$ operations are introduced because we know a priori that probabilities are bounded between $0$ and $1$. The assignment of an error $\epsilon^{p}_1$ to the empirical result $\hat{p}_1$ relies on a statistical analysis and depends on the employed statistical bound, such as Chebysev, Chernoff (Hoeffding) and Clopper-Pearson bounds.
Here one of the main differences with respect to other algorithms in the literature becomes obvious: although the probabilities are bounded between $0$ and $1$, the estimated amplitude obtained by the identity \eqref{eqn:first_empirical_amplitude_estimation} is now bounded between $-1\leq a_1\leq 1$. In other words, $\bm{a_1}$ \textbf{can be negative} (see Figure \ref{fig:angle_problem}). Note that the sign of the amplitude depends on the sign of $b_1$, which is taken positive for simplicity. However, this election is arbitrary and it could be chosen negative. 

\subsection{Following iterations: amplifying the probability and shrinking the interval}

\begin{figure}[hbtp!]
        \centering
    \begin{subfigure}[t]{0.3\textwidth}
        \centering
        \includegraphics[width=\textwidth]{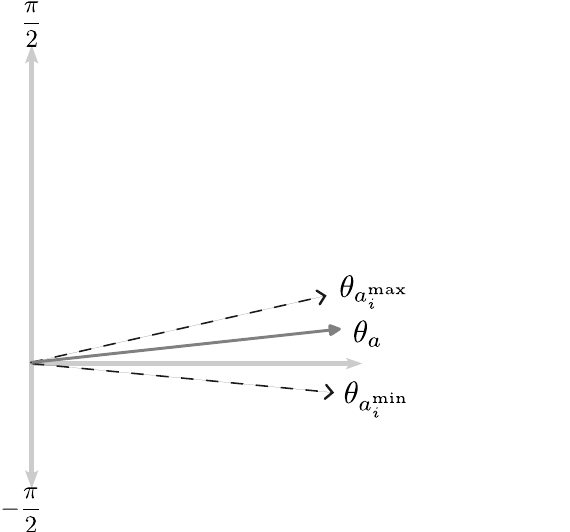}
        \caption{Starting point.}
        \label{fig:angle_problem_image2}
    \end{subfigure}
    \hfill
    \begin{subfigure}[t]{0.3\textwidth}
        \centering
        \includegraphics[width=\textwidth]{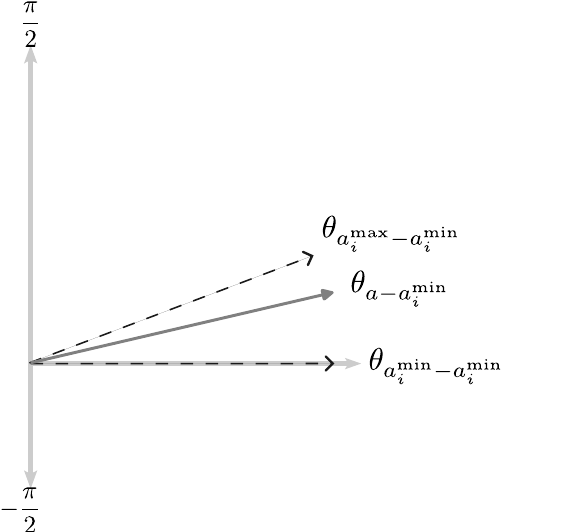}
        \caption{Shift.
        } \label{fig:shift}
    \end{subfigure}
    \hfill 
     \begin{subfigure}[t]{0.3\textwidth}
        \centering
        \includegraphics[width =\textwidth]{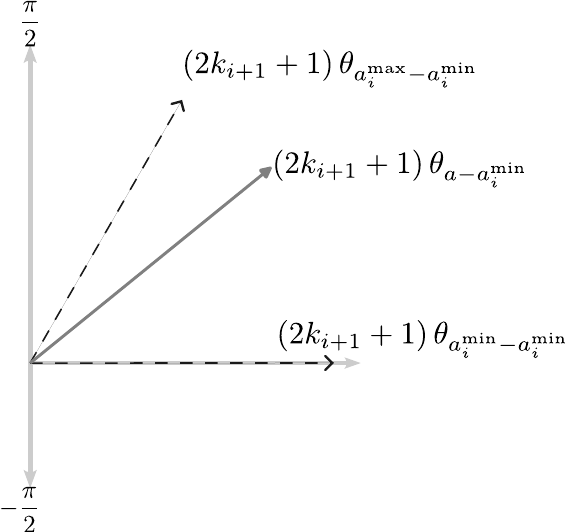}
        \subcaption{Amplification.
        }\label{fig:grover_operator}
    \end{subfigure}
        \\
     \begin{subfigure}[t]{0.3\textwidth}
        \centering
    \includegraphics[width=\textwidth]{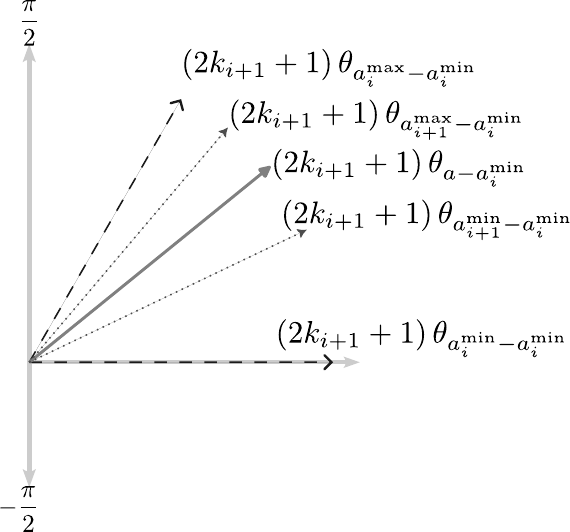}
    \subcaption{Measuring.
    }\label{fig:amplified_space_sampling}
    \end{subfigure}
\hfill
     \begin{subfigure}[t]{0.3\textwidth}
        \centering
        \includegraphics[width=\textwidth]{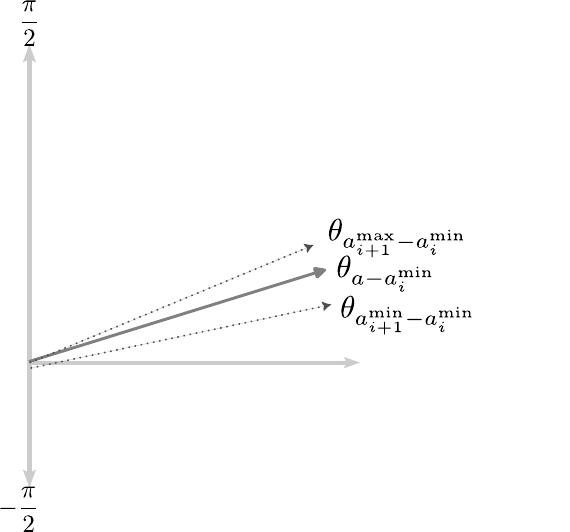}
         \subcaption{Undoing the amplification.
         }
    \label{fig:unamplified}
        \end{subfigure}
    \hfill
    \begin{subfigure}[t]{0.3\textwidth}
        \centering
        \includegraphics[width=\textwidth]{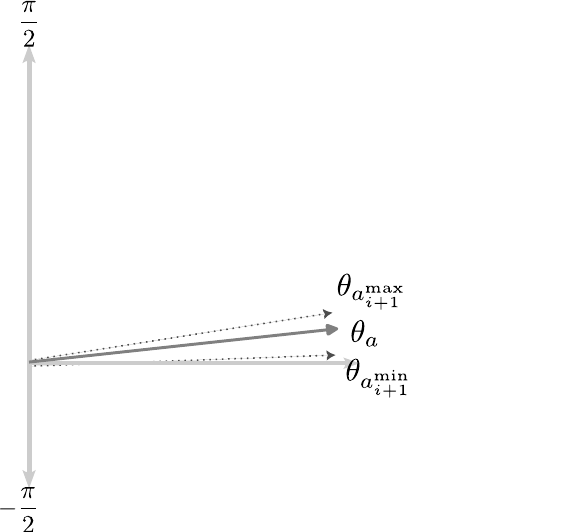}
        \subcaption{Undoing the shift.
        }\label{fig:second_iteration}
    \end{subfigure}
    \caption{ Graphical representation of the steps performed in the $i+1$ iteration. The solid grey line represents the unknown target value (which is not necessarily the center of the confidence interval). The dashed lines represent the bounds obtained at the $i$-th step, while the dotted lines represent the bounds obtained at step $i+1$. }
    \label{fig:following_iterations}
\end{figure}

On consecutive iterations, given an input confidence interval $(a^{\min}_i,a^{\max}_i)$ (see Figure \ref{fig:angle_problem_image2}) we want to obtain a tighter one $(a^{\max}_{i+1},a^{\min}_{i+1})$ and iterate the process until the desired precision $\epsilon$ is reached. At each iteration, the process for narrowing the interval starts by choosing a new shift according to 
\begin{equation}\label{shifti+1}
    b_{i+1} = -a^{\min}_{i}\ .
\end{equation}
Note that, this election is not unique, as we could have chosen  $b_{i+1} = -a^{\max}_{i}$ instead. Always keep in mind that the phase that we are obtaining is relative to the original value of $b = b_1$.
By considering the choice \eqref{shifti+1}, we force our lower bound to match exactly zero (see Figure \ref{fig:shift}). The boundaries of the confidence interval $(a^{\min}_i,a^{\max}_i)$, when shifted and then expressed in terms of the corresponding angles, become:
\begin{equation}\label{alphamax}
\begin{aligned}
    &\alpha^{\min}_{i} = 0,
    &\alpha^{\max}_{i} = \arcsin\left(a^{\max}_i-a^{\min}_i\right) = \arcsin\left(2\epsilon^a_i \right).   
\end{aligned}
\end{equation}
The angular region $\alpha_i^{\min} \leq \alpha_i \leq \alpha_i^{\max}$ represents the confidence interval and we refer to it as the confidence fan.\\ \\
The next step takes advantage of the Grover operator, defined as
\begin{equation}\label{groove}
    \mathcal{G} = -\mathcal{A}_{b}\mathcal{R}_{\ket{0}}\mathcal{A}_{b}^{\dagger}\mathcal{R}_{\ket{\phi}}\ ,
\end{equation}
where 
\begin{equation}
   \begin{aligned}
    &\mathcal{R}_{\ket{0}} &=\mathds{1}-2\ket{0}\bra{0}, \quad
    &\mathcal{R}_{\ket{\phi}} &=\mathds{1}-2\ket{\phi}\bra{\phi}\ ,
\end{aligned} 
\end{equation}
and $\mathcal{A}_{b}$ is the oracle defined in \eqref{eq:shi_def}.
The Grover operator applied $k_{i+1}$ times transforms the generic angle $\theta$ into $(2k_{i+1}+1)\theta$, see Figure \ref{fig:grover_operator}. Hence, the state $\ket{\psi_{i+1}}_+$ is transformed to:
\begin{equation}\label{eq:tet}
	\begin{aligned}
    \ket{\psi_{i+1}}_+ &= \left(a+b_{i+1}\right)\ket{\phi}+c_i\ket{\phi^\perp}= \left(a-a_i^{\min}\right)\ket{\phi}+c_i\ket{\phi^\perp}\\
    &\equiv \sin(\theta_{i+1})\ket{\phi}+\cos(\theta_{i+1})\ket{\phi^\perp}\\
    &\quad \xrightarrow{{\cal G}^{k_{i+1}}}\quad\\ 
    &\sin\left((2k_{i+1}+1)\theta_{i+1}\right)\ket{\phi}+\cos\left((2k_{i+1}+1)\theta_{i+1}\right)\ket{\phi^\perp}\ ,
\end{aligned}
\end{equation}
where the sub-index ``$+$" is employed as in \eqref{mas}, ${\cal G}^{k_{i+1}}$ indicates the Grover operator applied $k_{i+1}$ times and, in the second equality, we use \eqref{shifti+1}. Moreover, in \eqref{eq:tet} we have implicitly defined the angle $\theta_{i+1} \equiv \arcsin \left(a-a_i^{\text{min}}\right)$ for which we will use the analogous notation $\theta_{a-a_i^{\text{min}}}=\theta_{i+1}$, according to convenience of presentation.\\

\noindent
In order to avoid ambiguities due to the lack of a bijective correspondence angle/amplitude, when measuring amplified probabilities, we cannot allow the amplified angles to go beyond $\left[0,\dfrac{\pi}{2}\right]$. Namely, we need the amplified confidence fan to stay within the first quadrant. Relying on \eqref{alphamax}, we choose the Grover amplification exponent as:
\begin{equation}\label{eq:next_k}
k_{i+1} = \left\lfloor\dfrac{\pi}{4\arcsin\left(2\epsilon^a_{i}\right)}-\dfrac{1}{2}\right\rfloor\ ,
\end{equation}
so that we maximize the amplification factor while respecting the angle constraint.\\ \\

Next, we measure the state $\ket{\psi_{i+1}}$ in the amplified space, obtaining the empirical probability 
\begin{equation}
    \hat{p}_{i+1}\approx \sin^2 \left((2k_{i+1}+1)\theta_{i+1}\right)\ ,
\end{equation} 
with the statistical error $ \epsilon^{p}_{i+1}$, and define:
\begin{equation}\label{eq:pmaxmin}
    \begin{aligned}
    & p^{\min}_{i+1} := \max\left(\hat{p}_{i+1}-\epsilon^{p}_{i+1},0\right)\ ,
    & p^{\max}_{i+1} := \min\left(\hat{p}_{i+1}+\epsilon^{p}_{i+1},1\right)\ , 
    \end{aligned}
\end{equation}
and,
\begin{equation}
    p_{i+1} := \dfrac{p^{\max}_{i+1}+p^{\min}_{i+1}}{2}\ ,
\end{equation}
where the $\max$ and $\min$ functions play an analogous role as in Section \ref{sign} (see Figure \ref{fig:amplified_space_sampling}).\\ \\
In the next step we transform the angles corresponding to $p^{\max}_{i+1}$ and $p^{\min}_{i+1}$ to the non-amplified space:
\begin{equation}
    \begin{aligned}
    & \theta^{\min}_{i+1} =  \dfrac{\arcsin\left(\sqrt{p^{\min}_{i+1}}\right)}{2k_{i+1}+1},
    & \theta^{\max}_{i+1} = \dfrac{\arcsin\left(\sqrt{p^{\max}_{i+1}}\right)}{2k_{i+1}+1}\ .
\end{aligned}
\end{equation}
In other words, we have just ``undone" the amplification (see Figure \ref{fig:unamplified}). \\ \\

Finally, we have to undo the shift \eqref{shifti+1}, actually performing an opposite shift (see Figure \ref{fig:second_iteration}). Using definitions analogous to those given in \eqref{eq:first_amplitude_bounds}, we
finally obtain:
\begin{equation}
   \begin{aligned}\label{eqn:following_amplitude_bounds}
    & a^{\min}_{i+1} =  \sin\left(\dfrac{\arcsin\left(\sqrt{p^{\min}_{i+1}}\right)}{2k_{i+1}+1}\right)-b_{i+1}\ ,
    & a^{\max}_{i+1} = \sin\left(\dfrac{\arcsin\left(\sqrt{p^{\max}_{i+1}}\right)}{2k_{i+1}+1}\right)-b_{i+1}\ ,
    \end{aligned} 
\end{equation}
so that,
\begin{equation}
   \begin{aligned}
   & a_{i+1} = \dfrac{a^{\max}_{i+1}+a^{\min}_{i+1}}{2}\ ,\\
    & \epsilon^a_{i+1} = \dfrac{a^{\max}_{i+1}-a^{\min}_{i+1}}{2} = \frac{1}{2}\sin\left(\dfrac{\arcsin\left(\sqrt{p^{\max}_{i+1}}\right)}{2k_{i+1}+1}\right)-\frac{1}{2}\sin\left(\dfrac{\arcsin\left(\sqrt{p^{\min}_{i+1}}\right)}{2k_{i+1}+1}\right).
\end{aligned} 
\end{equation}

Recall that the goal is to reduce the width of the confidence interval until the desired precision $\epsilon$ is reached. For this purpose, one has to repeat the iteration just described until the goal is met.\\ \\


\begin{remark}\label{remark:mrqae}
Throughout this section we have addressed the general structure of the algorithm. Nevertheless, we have not specified the values of all the involved parameters. More specifically, we have not discussed how $\epsilon^p_i$ is obtained. This parameter strictly depends on the number of shots on each iteration, $N_i$, and the confidence required on each iteration, $1-\gamma_i$, through a set of bounds such as Hoeffding's inequality or Clopper-Pearson bound (see \cite{hoeffding,clopper_pearson}). In Section \ref{sec:configuration_properties}, more insight about these choices is provided.
\end{remark}

\subsection{Configuration and properties}\label{sec:configuration_properties}
    As mentioned in Remark \ref{remark:mrqae}, in order to complete the mRQAE method, we need to incorporate a particular choice for the parameters $\epsilon^p_i$ and $b_1$ and, thus, the parameters involved in its computation, \emph{i.e.}, $N_i$ and $1-\gamma_i$, the number of shots and confidence level of the $i$-th iteration, respectively. In the following theorem we describe a possible configuration, where there are three main parameters, $\epsilon$, $1-\gamma$ and $q$. The first one is the target precision, the second one the target confidence level and the third one is the amplification policy. The amplification policy is a lower bound on the ratio between the number of aplications of the Grover oracle between subsequent iterations.
    
\begin{theorem}\label{the:mrqae_properties}
	
	Given $\epsilon,\gamma$ and $q$, and taking the parameters:
	\begin{equation}\label{eq:mrqae_number_shots}
		N_i(q,\epsilon,\gamma) = \left\lceil \dfrac{1}{2\epsilon^p(q,k_i)^2}\log\left(\dfrac{2}{\gamma_i}\right) \right\rceil,
	\end{equation}
	\begin{equation}\label{eq:mrqae_max_confidence}
		\gamma_i(q,\epsilon,\gamma) = \dfrac{\gamma}{2} \dfrac{q-1}{q}\dfrac{2k_i+1}{2k^{\max}+1},
	\end{equation}
	\begin{equation}\label{eq:mrqae_b1}
		 b_1 = \dfrac{1}{2}\ ,
	\end{equation}
	with
	\begin{equation}\label{eq:mrqae_min_precision}
		\epsilon^p (q,k_i) = 
		\begin{cases}
		\dfrac{1}{2}\sin\left(\dfrac{\pi}{2(q+2)}\right), \quad \text{if}\quad k_i = 0\\ \dfrac{1}{2}\sin^2\left(\dfrac{\pi}{4\left(q+\dfrac{2}{2k_i+1}\right)}\right), \quad \text{if} \quad k_i >0
		\end{cases}\ ,
	\end{equation}
	\begin{equation}\label{eq:mrqae_max_iterations}
		 k^{\max}(q,\epsilon) = \left\lceil \dfrac{\arcsin\left(\sqrt{2\epsilon^p (q,\infty)}\right)}{\arcsin(2\epsilon)}-\dfrac{1}{2}\right\rceil\ ,
	\end{equation}
	then, we have the following properties:
	\begin{enumerate}[label=\roman*)]
		\item The error $\epsilon_i^p$ at iteration $i$ is bounded by:
		\begin{equation}\label{eq:mrqae_precision_bound}
			\epsilon^p_i \leq \epsilon^p(q,k_i)\ .
		\end{equation}
		\item The amplification policy $q_i$ is lower bounded at any iteration by:
		\begin{equation}\label{eq:mrqae_amplification_policy_bound}
			q_i = \dfrac{2k_{i+1}+1}{2k_i+1}\geq q\ .
		\end{equation}
		\item The maximum depth of the circuit $k_I$ is upper bounded by:
		\begin{equation}\label{eq:mrqae_depth_bound}
			k_I\leq k^{\max} (q,\epsilon).
		\end{equation}
		\item The maximum number of iterations $I$ is upper bounded by:
		\begin{equation}\label{eq:mrqae_max_iterations_bound}
			I<T = \log_q\left(q^2\dfrac{2\arcsin\left(\sqrt{2\epsilon^p(q,\infty)}\right)}{\arcsin\left(2\epsilon\right)}\right)>I.
		\end{equation}
		\item The algorithm obtains a precision $\epsilon$ with confidence $1-\gamma$ (Proof of Correctness):
		\begin{equation}\label{eq:mrqae_master_equation}
			\mathbb{P}\Big[a\not\in (a^{\min}_I,a^{\max}_I)\Big]\leq \gamma.
		\end{equation}
		
		\item The total number of calls to the oracle is bounded by:
		\begin{align}\label{eq:mrqae_oracle_bound}
			N_Q< C_1(q)\dfrac{1}{\epsilon}\log\left(\dfrac{C_2(q)}{\gamma}\right)\ ,
		\end{align}
		where $C_1(q),\, C_2(q)$ are two constants that depend on q.
	\end{enumerate}
\end{theorem}
\begin{proof}\\ \\
\textbf{\textit{Proof of property i)}}
When finding an empirical estimate $\hat{p}$ of a probability $p$, we can assign to it a confidence interval (\emph{i.e.} we can estimate an associated statistical error) by using Hoeffding's inequality\footnote{Although there exist tighter bounds, they are much less tractable from an analytic point of view. One such example is  Clopper-Pearson \cite{clopper_pearson}.} (see \cite{hoeffding}):%
\begin{equation}\label{eqn:mrqae_Hoeffding_bounds}
	\mathbb{P}\Big(\left|p-\hat{p}\right|\geq {\epsilon}_p \Big)\leq 2e^{-n \epsilon_p^2} = \gamma\ ,
\end{equation}
where $\epsilon_p$ is the precision, $1-\gamma$ is the confidence level and $n$ is the number of shots (\emph{i.e.} samplings) used for the measurement. As we fixed the values for $N_i$ and $\gamma_i$ in Equations \eqref{eq:mrqae_number_shots} and \eqref{eq:mrqae_max_confidence}, using \eqref{eqn:mrqae_Hoeffding_bounds} we get a fixed value for $\epsilon^p_i$:
\begin{equation}
	\mathbb{P}\Big[\left|\sin^2\left[(2k_i+1)\theta_i\right]-\hat p_i\right|\geq\epsilon^p_i\Big] \leq 2e^{-2N_i(\epsilon^p_i)^2} = \gamma_i\ .
\end{equation}
Rewriting the previous expression in terms of $\epsilon^p_i$ we have:
\begin{equation}
	\begin{aligned}
		\left(\epsilon^p_i\right)^2 &= \dfrac{1}{2N_i}\log\left(\dfrac{2}{\gamma_i}\right) = \\
		& = \dfrac{1}{2\left\lceil \dfrac{1}{2\epsilon^p(q,k_i)^2}\log\left(\dfrac{2}{\gamma_i}\right) \right\rceil}\log\left(\dfrac{2}{\gamma_i}\right)\\
        &\leq \dfrac{1}{2\left[\dfrac{1}{2\epsilon^p(q,k_i)^2}\log\left(\dfrac{2}{\gamma_i}\right)\right]}\log\left(\dfrac{2}{\gamma_i}\right) = (\epsilon^p(q,k_i))^2\ ,
	\end{aligned}
\end{equation}
where we used \eqref{eq:mrqae_number_shots}. We have thus proven property \textbf{\textit{i)}}.\\ \\
\textbf{\textit{Proof of property ii)}} We first consider the case $i>1$. From Equation \eqref{eq:next_k}, we have:
\begin{equation}
	q_i = \dfrac{2k_{i+1}+1}{2k_i+1} = \dfrac{2\left\lfloor\dfrac{\pi}{4\arcsin(2\epsilon^a_i)}-\dfrac{1}{2}\right\rfloor+1}{2k_i+1}\ .
\end{equation}
We now consider the fact that $\left\lfloor x\right\rfloor\geq x-1$ thus obtaining:
\begin{equation}\label{eq:mrqae_intermediate_step}
	q_i \geq \dfrac{\dfrac{\pi}{2\arcsin(2\epsilon^a_i)}-2}{2k_i+1} = \dfrac{\pi}{2\arcsin(2\epsilon^a_i)(2k_i+1)}-\dfrac{2}{2k_i+1}\ .
\end{equation}
Now we focus on the term $(2k_i+1)\arcsin(2\epsilon_i^a)$ which can be rewritten in terms of $\epsilon^p_i$ as
\begin{equation}\label{eq:mrqae_ampi_k}
	(2k_{i}+1)\arcsin\left[ \sin\left(\dfrac{\arcsin\left(\sqrt{\min(\hat{p}_i+\epsilon^p_i,1)}\right)}{2k_{i}+1}\right)-\sin\left(\dfrac{\arcsin\left(\sqrt{\max(\hat{p}_i-\epsilon^p_i,0)}\right)}{2k_{i}+1}\right)\right]\ ,
\end{equation}
where we have used \eqref{eqn:following_amplitude_bounds} and \eqref{eq:pmaxmin} after an obvious relabelling of the index. 

In what follows, we use some lemmas in Appendix \ref{ch:appendix1}. Using Lemma \ref{lemma:upper_bound_sine} we upper bound the expression (note that the values within the $sin$ functions range between $0$ and $1$):
\begin{equation}
	(2k_i+1)\arcsin(2\epsilon_i^a)\leq \arcsin\left(\sqrt{\min(\hat{p}_i+\epsilon^p_i,1)}\right)-\arcsin\left(\sqrt{\max(\hat{p}_i-\epsilon^p_i,0)}\right).
\end{equation}
Next, we upper bound the difference of arcsin functions using Lemma \ref{lemma:upper_bound_arcsin} with:
\begin{equation}
	(2k_i+1)\arcsin(2\epsilon_i^a)\leq 2\arcsin\left(\sqrt{\min(\hat{p}_i+\epsilon^p_i,1)}-\sqrt{\max(\hat{p}_i-\epsilon^p_i,0)}\right).
\end{equation}
Then, using Lemma \ref{lemma:upper_bound_square_root} we upper bound the difference of the square roots and get:
\begin{equation}
	(2k_i+1)\arcsin(2\epsilon_i^a)\leq 2\arcsin\left(\sqrt{\min(\hat{p}_i+\epsilon^p_i,1)-\max(\hat{p}_i-\epsilon^p_i,0)}\right).
\end{equation}
After that, we simply upper bound the max and min functions by their values:
\begin{equation}\label{eq:mrqae_some_equation}
	(2k_i+1)\arcsin(2\epsilon_i^a)\leq 2\arcsin\left(\sqrt{2\epsilon^p_i}\right).
\end{equation}
Finally, we can use Equation \eqref{eq:mrqae_some_equation} to bound the expression in Equation \eqref{eq:mrqae_intermediate_step}, \emph{i.e.},
\begin{equation}
	q_i \geq \dfrac{\pi}{2\arcsin(2\epsilon^a_i)(2k_i+1)}-2\geq \dfrac{\pi}{4\arcsin\left(\sqrt{2\epsilon^p_i}\right)}-\dfrac{2}{2k_i+1}\ .
\end{equation}
As $\epsilon^p_i \leq \epsilon^p(q,k_i)$, we have that:
\begin{equation}
	q_i\geq q\ .
\end{equation}
So far, we have not treated the first iteration, $i= 1$, which we now consider explicitly:
\begin{equation}\label{eq:mrqae_intermediate_first_iteration}
	q_1 = \dfrac{2k_2+1}{2k_1+1} = 2k_2+1 = 2\left\lfloor\dfrac{\pi}{4\arcsin(2\epsilon^a_1)}-\dfrac{1}{2}\right\rfloor+1 \geq \dfrac{\pi}{2\arcsin(2\epsilon^a_1)}-2\ ,
\end{equation}
where we have recalled that $k_1 = 0$.
We focus our attention on the term
\begin{equation}
	\arcsin(2\epsilon^a_1) = \arcsin\left[ \min\left(\dfrac{\hat{p}_{\text{sum}}-\hat{p}_{\text{diff}}}{4b_1}+\dfrac{\epsilon^{p}_1}{|2b_1|},1\right)
	-\max\left(\dfrac{\hat{p}_{\text{sum}}-\hat{p}_{\text{diff}}}{4b_1}-\dfrac{\epsilon^{p}_1}{|2b_1|},-1\right)\right]\ ,
\end{equation}
where we have considered \eqref{eq:first_amplitude_bounds}. 
Following the same strategy as before we define:
\begin{equation}\label{eq:mrqae_b_function}
	f(\epsilon^p_1) = \arcsin\left[ \min\left(\dfrac{\hat{p}_{\text{sum}}-\hat{p}_{\text{diff}}}{4b_1}+\dfrac{\epsilon^{p}_1}{|2b_1|},1\right)
	-\max\left(\dfrac{\hat{p}_{\text{sum}}-\hat{p}_{\text{diff}}}{4b_1}-\dfrac{\epsilon^{p}_1}{|2b_1|},-1\right)\right]\ .
\end{equation}
An upper bound for \eqref{eq:mrqae_b_function} is:
\begin{equation}\label{sec:mrqae_b_function_upper}
	\overline{f}(\epsilon^p_1) = \arcsin\left( \dfrac{\epsilon^p_1}{\left|b_1\right|}\right)\geq f(\epsilon_1^p)\ ,
\end{equation}
which can be obtained by using Lemma \ref{lemma:upper_bound_arcsin}.
Hence, from Equation \eqref{eq:mrqae_intermediate_first_iteration}, we have that:
\begin{equation}\label{eq:mrqae:dise2}
	q_1 \geq \dfrac{\pi}{2\arcsin\left(\dfrac{\epsilon^p}{|b_1|}\right)}-2\ .
\end{equation}

Eventually, by the definition of $b_1$ we have:
\begin{equation}
	q_1\geq q\ ,
\end{equation}
and we have proven proposition \textbf{\textit{ii)}}.\\ \\
\textbf{\textit{Proof of property iii)}} Using \eqref{eqn:following_amplitude_bounds}, if we get the maximum $k$, $k_{\max}$ we have
\begin{equation}
	\epsilon^a_I(k^{\max}) = \frac{1}{2}\sin\left(\dfrac{\arcsin\left(\sqrt{p^{\max}_{I}}\right)}{2k^{\max}+1}\right)-\frac{1}{2}\sin\left(\dfrac{\arcsin\left(\sqrt{p^{\min}_{I}}\right)}{2k^{\max}+1}\right)\ .
\end{equation}
Using Lemmas \ref{lemma:upper_bound_sine}, \ref{lemma:upper_bound_arcsin} and \ref{lemma:upper_bound_square_root}, we obtain:
\begin{equation}
	\epsilon^a_I(k^{\max}) \leq \dfrac{1}{2}\text{sin}\left(\dfrac{2\text{arcsin}\left(\sqrt{2\epsilon^p_i}\right)}{2k^{\max}+1}\right)\leq  \dfrac{1}{2}\text{sin}\left(\dfrac{2\text{arcsin}\left(\sqrt{2\epsilon^p(q,k^{\max})}\right)}{2\left\lceil\dfrac{\arcsin\left(\sqrt{2\epsilon^p(q,\infty)}\right)}{\arcsin(2\epsilon)}-\dfrac{1}{2}\right\rceil+1}\right) \leq \epsilon\ ,
\end{equation}
where we have also used property \textit{\textbf{i)}} and the definition of $k^{\max}$. So, we have proven property \textbf{\textit{iii)}}.\\ \\
\textbf{\textit{Proof of property iv)}} In this subsection we bound the maximum number of iterations needed to achieve the target accuracy $\epsilon$. First note that, if $I$ represents the last iteration, we have that
\begin{equation}\label{eq:mrqae_penul}
	\epsilon<\epsilon^a_{I-1} =  \frac{1}{2}\sin\left(\dfrac{\arcsin\left(\sqrt{p^{\max}_{I-1}}\right)}{2k_{I-1}+1}\right)-\frac{1}{2}\sin\left(\dfrac{\arcsin\left(\sqrt{p^{\min}_{I-1}}\right)}{2k_{I-1}+1}\right),
\end{equation}
otherwise we would be in the last iteration, and that is false by hypothesis. In order to write \eqref{eq:mrqae_penul} we have used \eqref{eqn:following_amplitude_bounds} with $i=I-1$. Using similar arguments as in property \textbf{\textit{iii}}, we bound $\epsilon^a_{I-1}$ by
\begin{equation}\label{eq:mrqae_target_bou}
	\epsilon < \epsilon^a_{I-1} \leq \frac{1}{2}\sin\left(\dfrac{2\arcsin\left(\sqrt{2\epsilon^p_{I-1}}\right)}{2k_{I-1}+1}\right)\ .
\end{equation}
So, we can rewrite \eqref{eq:mrqae_target_bou} as
\begin{equation}\label{eq:mrqae_rt2}
	\begin{aligned}
		&(2k_1+1)\prod_{i = 1}^{I-2} q_i = 2k_{I-1}+1 <\dfrac{2\arcsin\left(\sqrt{2\epsilon^p_{I-1}}\right)}{\arcsin(2\epsilon)}\\
		&\leq \dfrac{2\arcsin\left(\sqrt{2\epsilon^p(q,\infty)}\right)}{\arcsin(2\epsilon)} =: (2k_1+1)\,\prod_{i = 1}^{T-2}q = (2k_1+1) q^{T-2}\ ,
	\end{aligned}
\end{equation}
where we have used $\epsilon^p_{i}\leq \epsilon^p(q,\infty)$ and we have introduced the positive number $T$. Additionally, from \eqref{eq:mrqae_rt2}, we obtain that
\begin{equation}\label{eq:mrqae_max_iterations_aux}
	\prod_{i = 1}^{I-2} q_i<q^{T-2}\ .
\end{equation}
By using property \textit{\textbf{ii)}}, we get
\begin{equation}
	T = \log_q\left(q^2\dfrac{2\arcsin\left(\sqrt{2\epsilon^p(q,\infty)}\right)}{\arcsin(2\epsilon)}\right)>I,
\end{equation}\label{eq:upper_bound_T}
thus proving property \textbf{\textit{iv)}}.
\textbf{\textit{Proof of property v)}} We aim to ensure that the precision $\epsilon$ is met with confidence $1-\gamma$. In order to achieve this, note that:
\begin{align*}
	\mathbb{P}\Big[a\not\in [a^{\min}_I,a^{\max}_I]\Big] &=\mathbb{P}\Big[\sin^2\left[(2k_I+1)\theta_{I}\right]\not\in [p^{\min}_I,p^{\max}_I]\Big]\\
	&\leq \mathbb{P}\Big[\bigcup\limits_{i = 1}^{I} \sin^2\left[(2k_i+1)\theta_{i}\right]\not\in [p^{\min}_i,p^{\max}_i]\Big]\\
	&\leq \sum\limits_{i = 1}^I \mathbb{P}\Big[\sin^2\left[(2k_i+1)\theta_{i}\right]\not\in [p^{\min}_i,p^{\max}_i]\Big] \\ 
	&\leq \sum\limits_{i = 1}^I\gamma_i= \dfrac{\gamma}{2}\dfrac{1}{2k^{\max}+1}\dfrac{q-1}{q}+
	\dfrac{\gamma}{2}\dfrac{1}{2k^{\max}+1}\dfrac{q-1}{q}\sum\limits_{i = 2}^I 2 k_i+1,
\end{align*}
where we have used the definition of $\gamma_i$ and Boole's inequality. 

Next, applying Lemma \ref{lemma:upper_bound_series} with $f(x) = x$ and $2k_{i+1}+1\geq q (2k_i+1)$, we obtain:
\begin{align*}
	\mathbb{P}\Big[a\not\in [a^{\min}_I,a^{\max}_I]\Big]&\leq 
	\dfrac{\gamma}{2}\dfrac{1}{2k^{\max}+1}\dfrac{q-1}{q}+\dfrac{\gamma}{2}\dfrac{q-1}{q}\dfrac{1}{2k^{\max}+1}\sum\limits_{i = 1}^{I-1} \dfrac{2k^{\max}+1}{q^i} \\
 &\leq \gamma \dfrac{q-1}{q}\sum\limits_{i = 0}^{\infty} \dfrac{1}{q^i} = \gamma,
\end{align*}
where we have used the property \textbf{\textit{iii)}}. We recall that $1-\gamma_i$ represents the confidence level of the $i$-th iteration.\\ \\
\textbf{\textit{Proof of property vi)}} We want to find an upper bound on the number of calls to the Grover oracle in order to obtain a target precision $\epsilon$ with confidence $1-\gamma$.

As we finish after $I$ iterations, the number of calls to the Grover oracle is given by
\begin{equation}\label{eq:calls_oracle1}
	N_Q = \sum_{i = 1}^{I}N_ik_i < \dfrac{1}{2}\sum_{i = 1}^{I}N_i(2k_i+1). 
\end{equation}
Next, we need an upper bound $\overline{N}_i$ for the number of shots $N_i$ at each iteration:
\begin{equation}\label{eq:upper_bound_shots}
	N_i = \left\lceil \dfrac{1}{2\epsilon^p(q,k_i)^2}\log\left(\dfrac{2}{\gamma_i}\right) \right\rceil< \dfrac{1}{2\epsilon^p(q,k_i)^2}\log\left(\dfrac{2}{\gamma_i}\right)+1<\dfrac{1}{2\epsilon^p(q,0)^2}\log\left(\dfrac{2\sqrt{e}}{\gamma_i}\right)=:\overline{N}_i,
\end{equation}
where we have redefined $\epsilon^p(q,0)$ with respect to Equation \eqref{eq:mrqae_min_precision} as:
\begin{equation}
    \epsilon^p(q,0) = \dfrac{1}{2}\sin^2\left(\dfrac{\pi}{4(q+2)}\right).
\end{equation}
Replacing \eqref{eq:upper_bound_shots} in Equation \eqref{eq:calls_oracle1} we get:
\begin{equation*}
	\begin{aligned}
	N_Q &< \dfrac{1}{2}\sum_{i = 1 }^{I}\overline{N}_i(2k_i+1) = \dfrac{1}{4\epsilon^p(q,0)^2}\sum_{i = 1 }^{I}\log\left(\dfrac{2\sqrt{e}}{\gamma_i}\right)(2k_i+1) \\
	 & = \dfrac{1}{4\epsilon^p(q,0)^2}\sum_{i = 1 }^{I}\log\left(\dfrac{4\sqrt{e}q (2k^{\max}+1)}{\gamma (q-1) (2k_i+1)}\right)(2k_i+1). 
	\end{aligned}
\end{equation*}
Next, by using Lemma \ref{lemma:upper_bound_series} on the function $x\,\log\left(\dfrac{c}{x}\right)$ we can find an upper bound of the sum:
\begin{equation*}
	\begin{aligned}
		N_Q &<  \dfrac{1}{4\epsilon^p(q,0)^2}\sum_{i = 1 }^{I-1}\log\left(\dfrac{4\sqrt{e} q (2k^{\max}+1)q^i}{\gamma (q-1) (2k^{\max}+1)}\right)\dfrac{(2k^{\max}+1)}{q^i} \\
		&= \dfrac{2k^{\max}+1}{4\epsilon^p(q,0)^2}\sum_{i = 1 }^{I-1}\log\left(\dfrac{4\sqrt{e} q^{i+1}}{\gamma (q-1) }\right)\dfrac{1}{q^i}\\
		&=\dfrac{2k^{\max}+1}{4\epsilon^p(q,0)^2}\left(\log\left(\dfrac{4\sqrt{e} q}{\gamma (q-1) }\right)\sum_{i = 1 }^{I-1}\dfrac{1}{q^i}+\sum_{i = 1 }^{I-1}\log\left(q^i\right)\dfrac{1}{q^i}\right)\\
		&=\dfrac{2k^{\max}+1}{4\epsilon^p(q,0)^2}\left(\log\left(\dfrac{4\sqrt{e} q}{\gamma (q-1) }\right)\sum_{i = 1 }^{I-1}\dfrac{1}{q^i}+\log\left(q\right)\sum_{i = 1 }^{I-1}i\dfrac{1}{q^i}\right).
  	\end{aligned}
\end{equation*}
Then, in order to upper bound the sums we extend the summation of positive numbers up to infinity:
\begin{equation*}
	\begin{aligned}
		N_Q &< \dfrac{2k^{\max}+1}{4\epsilon^p(q,0)^2}\left(\log\left(\dfrac{4\sqrt{e} q}{\gamma (q-1) }\right)\sum_{i = 1 }^{I-1}\dfrac{1}{q^i}+\log\left(q\right)\sum_{i = 1 }^{I-1}i\dfrac{1}{q^i}\right)\\
	&\leq\dfrac{2k^{\max}+1}{4\epsilon^p(q,0)^2}\left(\log\left(\dfrac{4\sqrt{e} q}{\gamma (q-1) }\right)\sum_{i = 0 }^{\infty}\dfrac{1}{q^i}+\log\left(q\right)\sum_{i = 0 }^{\infty}i\dfrac{1}{q^i}\right)\\
  &\leq \dfrac{2k^{\max}+1}{4\epsilon^p(q,0)^2}\left(\log\left(\dfrac{4\sqrt{e} q}{\gamma (q-1) }\right)\dfrac{q}{q-1}+\log\left(q\right)\dfrac{q}{(q-1)^2}\right)\\
		& = \dfrac{2k^{\max}+1}{4\epsilon^p(q,0)^2}\dfrac{q}{q-1}\log\left(\dfrac{4\sqrt{e} q^{\frac{q}{q-1}}}{\gamma (q-1) }\right)\leq\\
		&\leq \left( 2\dfrac{\arcsin\left(\sqrt{2\epsilon^p(q,\infty)}\right)}{\arcsin(2\epsilon)}+2\right)\dfrac{1}{4\epsilon^p(q,0)^2}\dfrac{q}{q-1}\log\left(\dfrac{4\sqrt{e} q^{\frac{q}{q-1}}}{\gamma (q-1) }\right)\\
		&\leq \left( \dfrac{1}{\arcsin(2\epsilon)}\right)\dfrac{\pi}{2\epsilon^p(q,0)^2}\dfrac{q}{q-1}\log\left(\dfrac{4\sqrt{e} q^{\frac{q}{q-1}}}{\gamma (q-1) }\right)\\
		& \leq C_1(q)\dfrac{1}{\epsilon}\log\left(\dfrac{C_2(q)}{\gamma}\right),
	\end{aligned}
\end{equation*}
and we end the proof of property \textbf{\textit{vi)}}.
\end{proof}